\documentclass[aps,prl,twocolumn]{revtex4}
\usepackage{graphicx}
\usepackage{float}
\usepackage{setspace}
\usepackage{amsmath}

\begin{document}

\title{Another Look at Confidence Intervals: \\
Proposal for a More Relevant and Transparent Approach 
}  

\author{Steven D. Biller \\ {\it Department of Physics, University of Oxford, Oxford OX1 3RH, UK} \\
Scott M. Oser \\ {\it Department of Physics \& Astronomy, University of
  British Columbia, Vancouver V6T 1Z1, Canada} }




\begin{abstract}

The behaviors of various confidence/credible interval constructions
are explored, particularly in the region of low event numbers where
methods diverge most. We highlight a number of challenges, 
such as the treatment of nuisance parameters, and common misconceptions 
associated with such constructions.  An informal survey
of the literature suggests that confidence intervals are not always
defined in relevant ways and are too often misinterpreted and/or
misapplied. This can lead to seemingly paradoxical behaviors and
flawed comparisons regarding the relevance of experimental
results. We therefore conclude that there is a need for a more pragmatic
strategy which recognizes that, while it is critical to objectively
convey the information content of the data, there is also a strong
desire to derive bounds on model parameter values and a natural instinct to interpret
things this way. Accordingly, we attempt to put aside philosophical biases in favor of a  
practical view to propose a more transparent and 
self-consistent approach that better addresses these issues.

\end{abstract}


\maketitle


\section{Introduction}
The ability to distill experimental results in a form relevant to
theoretical models is fundamental to scientific inquiry. Yet the best
approach for this is still a matter of considerable discussion and
debate. At the heart of the issue is the desire to both objectively
quantify results in a frequentist manner and also
draw relevant inferences for specific models, which inherently
requires a Bayesian context ({\em i.e.} a choice of prior) for those
models. A failure to satisfactorily address both of these aspects has,
in many cases, led to misinterpretation and misapplication that have
not been mitigated by the adoption of new frequentist conventions. The
impact is largest for experiments working in the region of low numbers of signal
events, where different approaches diverge most. The confusion is
not helped by the use of forms for the display of frequentist
information that seem to suggest direct bounds on model parameter
values or relative experimental sensitivities to such models, neither
of which is necessarily the case. Suggestions that such confusion
arises from questions that should not be asked concerning
models are not satisfactory and fail to confront the fact that
scientists do, in fact, ask such questions
and should therefore make use of the appropriate formalism for these.

In fact, the goals of both objectively conveying the relevant information
content of data and deriving bounds on model parameter values are not mutually
exclusive, but rather are closely linked. It is not generally possible to 
translate experimental results into meaningful model constraints
without specifying a prior.
As such, detailed objective information should be
used to clearly define the context for Bayesian constraints.  The
issue is therefore largely one of establishing relevance and
transparency.

In this paper, we briefly review the nature of various interval
constructions; highlight some apparent paradoxes that arise from
common misinterpretations; cite specific cases where experiments have run into such issues; discuss several aspects associated with
practical implementation; and, finally, propose an approach to directly
address the above issues in a more relevant, self-consistent and
transparent manner using standard techniques.

\section{Interval Constructions and Their Meaning}

\subsection{Bayesian}
Bayesian probabilities quantify the degree of belief in a hypothesis. Given a
measurement, the goal of a Bayesian approach is to assign probabilities
to the range of possible model parameter values. By necessity, this requires an assumed
context for these models (prior), as indicated by Bayes' Theorem:

\begin{equation}
P(H_i|D) = \frac{P(D|H_i) P(H_i)}{\sum\limits_j P(D|H_j)P(H_j)}
\label{Bayes}
\end{equation}

\noindent where $P(H_i|D)$ is the posterior probability of hypothesis
$H_i$ given the data $D$; $P(D|H_i)$ is the likelihood of the data
assuming hypothesis $H_i$; and $P(H_i)$ is the prior probability for
$H_i$ that defines the {\em a priori} context relative to other
model parameter values. The ratio between Bayesian probabilities therefore provides an estimate of relative ``betting odds'' for which hypotheses are most likely to be correct.

For a purely Bayesian approach, there is no relevance of the concept of
``statistical coverage'' of a credible interval (the frequency with which 
a large number of repetitions of an experiment subject to random fluctuations 
would yield intervals that bound the correct hypothesis), since no comparison is done to a 
hypothetical ensemble ---  only the actual measurements matter. If desired, the effective statistical coverage can often
still be estimated for Bayesian constructions using Monte Carlo calculations etc. (as shown in Appendix A), 
but the credibility level that defines the construction simply relates the actual observation directly to the model.

Bayesian credible intervals are simply defined by the relevant
portion of the posterior probability density function (PDF) that
constitutes a fraction equal to a pre-defined credibility for the interval,
$CI$. The way this fraction is selected may be altered to
yield lower bounds, upper bounds, central intervals, the most compact
interval, or intervals containing the highest probability densities.
For intervals, as opposed to bounds, we suggest that using the highest
probability density offers the most intuitive and robust definition
for an arbitrary probability distribution.

As a simple example, we give the construction for an upper bound ({\em
i.e.} the critical value up to which integration is performed) on an
average signal strength, $S$, in a Poisson counting experiment where
the expected background level is $B$ and a total of $n$ events is
observed:

\begin{equation}
\frac{\int_{0}^{S_{up}}  [(S+B)^n e^{-(S+B)}/n!]P(S)dS}{\int_{0}^{\infty}[(S+B)^n e^{-(S+B)}/n!]P(S) dS}  = CI
\label{BayesPoisson}
\end{equation}

\noindent where $S_{up}$ is the upper bound to be determined, $P(S)$
is the prior probability for $S$, and $CI$ is the desired credibility
for the interval. In the case where all positive values of $S$ are {\em a priori}
given equal consideration ({\em i.e.} a uniform prior in which $P(S)$ is a constant for $S\ge0$), this can be
shown, by repeated integrations by parts, to be equivalent to:

\begin{equation}
\frac{\sum_{m=0}^n (S_{up}+B)^m e^{-(S_{up}+B)}/m!}{\sum_{m=0}^n B^m e^{-B}/m!} = 1-CI.
\label{BayesPoisson2}
\end{equation}

Thus, $S_{up}$ can be interpreted as denoting the upper limit on the
range of model parameter values for which the probability of observing $n$ events or
less is not more than $1-CI$, given that the possible number of
background events cannot be greater than the total number of events
observed in this measurement. If a non-uniform prior were used instead,
the form would be modified and the interpretation would be that the
upper limit is on the correspondingly weighted range of model parameter values.

\enlargethispage{\baselineskip}

\subsection{Standard Frequentist}

Frequentist probabilities are defined as the relative frequencies of occurrence
given a hypothetical ensemble of similar experiments subject to random fluctuations. 
There is no such thing as a ``probability'' for a model parameter to lie within
derived bounds --- either it does or it does not. However, if everyone
derived bounds in the same way, the correct model would be correctly bounded a
known fraction of the time (for more on statistical coverage, see Appendix A).

Rather than using the posterior probability, the Neyman construction of frequentist 
intervals~\cite{Neyman} starts with the probability density function (PDF) for a 
given observation under a fixed hypothesis that is used to construct the likelihood. 
For each possible
hypothesis, a portion of the possible outcomes containing the fraction $CL$ (frequentist confidence level)
is defined. The range of model parameter values for which a given measurement is
``likely'' ({\em i.e.} would be contained within that CL fraction)
then defines the confidence region. Note that this is not the same as
a statement that any given model is likely (which is Bayesian) and,
indeed, the construction is such as to avoid any direct comparison of
models. However, as before, there is an ambiguity in this construction
regarding how the PDF is used to compose the initial frequency
intervals, with common ordering choices including central, highest
probability density and most compact intervals. We will define
frequentist approaches that use an ordering principle based on the expected frequency of 
observations for a given hypothesis as ``standard frequentist." Approaches that fall outside
of this include those that use a likelihood ratio test as an alternative
ordering principle, such as Feldman-Cousins~\cite{fc} (which will be
discussed separately in the next section).

For comparison, the standard frequentist construction for an upper bound on an average
signal strength, $S$, in a Poisson counting experiment where the
expected background level is $B$ and a total of $n$ events is
observed can be written as follows:

\begin{equation}
\sum_{m=0}^n (S_{up}+B)^m e^{-(S_{up}+B)}/m! = 1-CL
\label{eq:Standard_Poisson}
\end{equation}

\vskip 0.1in

\noindent where $S_{up}$ can thus be interpreted as denoting the upper
limit on the range of model parameter values for which $n$ events or less would be
observed with a relative frequency of not more than $1-CL$ if the measurements
were to be repeated a large number of times. Note that this differs
from the Bayesian formula for a uniform prior only in the absence of the background
normalization. In other words, for this construction, the possible
number of background events is 
not constrained to be less than or equal to 
the total number of
all events observed in this particular measurement. 
This is because the probability being calculated is that
for observing $n$ events during a generic trial for an ensemble of
measurements, and does not take into account additional information
available from any particular observation (such as the fact that the
number of background events actually detected cannot exceed $n$).
Thus, the probability
associated with any particular measurement is not a meaningful concept
in the frequentist approach.

This can also be seen by the fact that the lack of a background
normalization means that there will be cases for which Equation
\ref{eq:Standard_Poisson} does not yield a positive solution for
$S_{up}$. These are instances where the observed number of events is
already deemed to be less probable than the desired confidence
level. Such ``empty intervals" are perfectly allowed and, indeed, are
necessary in order to guarantee the correct statistical coverage for
the frequency of observations within the overall ensemble of
hypothetical experiments. Individual frequentist bounds, however, do
not have meaning for model parameter values by themselves.  Indeed, for a case where
the confidence interval is empty, the observer knows that {\em for
this particular data set} the confidence interval does not contain the
true value of the parameter, even if the repeated construction of such confidence intervals would
correctly bound it in, say, 90\% of the cases where statistical fluctuations
resulted in different data sets. This distinction is fundamental: frequentist confidence
intervals are {\em always} statements about how often a large ensemble
of hypothetical experiments will bound the true value, and are {\em never} a
statement that there is a particular probability that the true value
is contained in the interval for any individual data set.  In fact, in
many cases for both standard frequentist and Feldman-Cousins intervals, the
experimenters may know that it is very unlikely that the true
model is contained in the generated interval for their particular data
set. This situation often tends to conflict with the desired interpretations of these bounds, since
the question of interest to most experimenters is the relevance of their own 
particular data set for the model parameter values under study, rather than the behavior of a large ensemble of hypothetical experiments that were not actually performed.

\subsection{Feldman-Cousins}

The approach of Feldman and Cousins~\cite{fc} uses an ordering principle for the
Neyman construction based on the ratio of likelihoods which, for the measurement 
of a quantity typified by a mean expectation, $\mu$, is given by:

\begin{equation}
\Lambda_{\mu}(x) = \frac{L(x|\mu)}{L(x|\mu_{best})}
\end{equation}

\noindent where $x$ is the measurement and $\mu_{best}$ is the mean
for the hypothesis in the physical region for which the data is most likely (not necessarily
the most likely hypothesis, an assessment of which would call for a
Bayesian construction).

\enlargethispage{\baselineskip}

In the standard frequentist case, the composition of intervals is simply based on the expected relative frequency of observations under each hypothesis. However, under the Feldman-Cousins approach, the composition of intervals is instead determined by a likelihood comparison across potentially different hypotheses, which can therefore lead to less intuitive interval choices.

\begin{figure}[H]
\includegraphics[width=85mm]{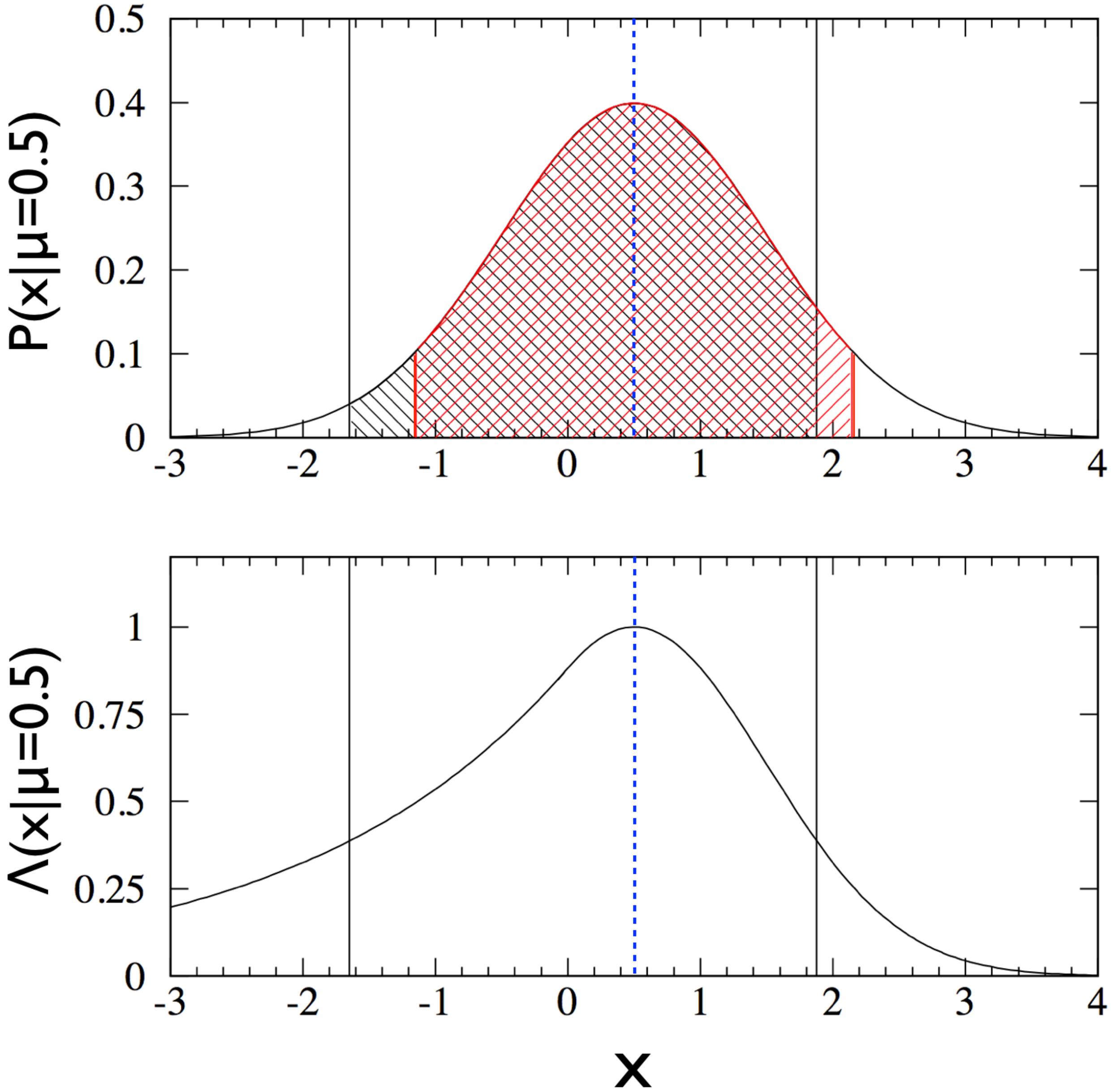}
\caption{Upper plot: Relative frequency for observations, $x$, of a Gaussian variable with $\sigma=1$ and $\mu=0.5$.
The range of observations for which the true mean is bounded for standard frequentist (red stripes) and Feldman-Cousins intervals (black stripes) at 90\% CL are shown. Lower plot: Values of the corresponding ordering parameter used to define the composition of the Feldman-Cousins interval.}
\label{Gauss_Bounds}
\end{figure}

As an example of how these interval definitions can differ, consider the case of a Gaussian variable with unit variance and a mean of $\mu=$0.5. Assume that this value of $\mu$ is unknown to us, but represents a physical quantity (such as a mass) that must be non-negative. From a given observation, we then wish to define 90\% CL bounds for $\mu$. Figure \ref{Gauss_Bounds} shows the relative frequency of observations, $P(x)$. The red striped area indicates the range of observations for which the correct value of $\mu$ is bounded by a central standard frequentist interval. These bounds are symmetric, extending $\pm1.65 \sigma$ relative to the value of $x=0.5$, as might be expected. However, as indicated by the black striped area, the range of observations for which the correct value of $\mu$ is bounded by a Feldman-Cousins frequentist interval is notably asymmetric (due to the fact that $\mu_{best}\ge 0$). The actual mean is included in the 90\% CL interval for an observation of $x=-1.6$, even though the observation is more than $2\sigma$ away from the true value. On the other hand, the interval excludes the true mean for an observation of $x=1.9$, even though this is less than $1.5\sigma$ away from the true mean and, hence, nearly 3 times more likely to occur. 

\vskip 0.1in

This counterintuitive result illustrates that the interpretation of
the Feldman-Cousins ordering principle, which is not based directly on
the frequency of the observation but instead on the likelihood ratio
$\Lambda(x|\mu)$, is not straightforward.

Feldman and Cousins cite Section 23.1 of the
5th edition of Kendall's book~\cite{kendall} as implying the ordering principle for
their interval construction, but, in fact, Sections 31.31-31.34 of that same reference are much more explicit regarding the general use of a likelihood ratio to define confidence intervals. These sections 
end with the statement, {\em ``The difficulties with such an approach are, as
before, the lack of a frequency interpretation for $P^*$ or, indeed,
any direct interpretation for the function. Here, as elsewhere, the statistician must decide
whether he or she is willing to make the logical leap in order to justify inferential statements
that relate to single experiments."}

One stated purpose of the Feldman-Cousins construction is to avoid empty intervals,
thus making the bounds appear more physical for the model. However, as
previously indicated, such empty intervals do not actually pose any problems
in principle since frequentist bounds do not refer to direct
restrictions on the physical model and only take on meaning for a
large ensemble of measurements, where statistical coverage is indeed
upheld. While the ordering principle used in the Feldman-Cousins
approach ensures that intervals are never empty, which many may find
less disconcerting, this does not avoid the basic issue that the bound
for any particular data set may not be meaningful, and
situations in which conventional frequentist intervals are empty are
often situations in which the Feldman-Cousins procedure returns a
value that is prone to misinterpretation as being an unduly strict bound on a model. 

Feldman and Cousins themselves recognized
this problem, noting that this results from a confusion with Bayesian inferences, which are more relevant for decision making. Accordingly, they recommended accompanying each limit by the
average expected limit (the ``sensitivity'' of the measurement) ``in order to
provide information that will assist in this (Bayesian) assessment.''
Cases in which the obtained limit is significantly better than the expected
sensitivity are cases where there is a higher probability that the true parameter 
value is not contained within the derived bounds. However, the expected limit 
clearly does not represent an actual measurement, and how information from this and/or the 
derived bounds is to be quantitatively applied in order to arrive at a Bayesian (or any other) assessment
for these cases is not at all clear.
Furthermore, this issue does not, in fact, have a clear threshold, as {\em all} 
negative fluctuations yield tighter bounds than the expected limit. Even worse, 
merely producing non-empty intervals by itself is not obviously an
improvement in relevance or clarity, since any particular interval (empty or not) constructed
for a given data set will not generally contain the true parameter value with a probability indicated
by the quoted confidence interval, in spite of a nearly
universal tendency to misinterpret it as such. The mere fact that
Feldman-Cousins returns non-empty intervals in some cases may actually
obfuscate their nature. In many ways a null interval, while
disconcerting, at least is transparently not a limit on the parameter,
whereas a  narrow but non-empty Feldman-Cousins interval,
such as often occurs when the data fluctuates below the expected
background, may give the false impression that the confidence interval
is meaningful for bounding a model.

\section{Examples of the Behavior of Upper Bounds for Low Numbers of Events}

We will now explore a number of scenarios in the region of low
event numbers that highlight the differences between various interval
constructions.

\vskip 0.2in

As an initial example, consider the scenario where the expected number
of background events, $B$, for 1 year of running with a 1 kg detector
is 9, but a statistical fluctuation results in a total of only 5
events observed. A fluctuation this low or lower would happen nearly
12\% of the time if the true signal rate were zero.
We now wish to construct a
90\% CL (frequentist) or CI (Bayesian)
upper bound, $S_{up}$, on a signal of
strength $S$.

\subsection{Standard Frequentist}

For the Standard Frequentist case, the above scenario yields from
Equation~\ref{eq:Standard_Poisson} a value of $S_{up}$ = 0.27 events
per kg per year.

Now consider the further scenario in which we were contemplating
running the experiment for an additional year. The second year of data
would very likely yield a number of background events much closer to
the expected mean, so we would likely end up with a total of
approximately 5+9=14 events where 18 are expected over the 2 year
run. The same formalism would then result in a value of $S_{up}$ =
2.12 events per 2 kg-years, or a bound on the rate of 1.06 events per kg
per year, which is nearly 4 times less restrictive than the limit from
the first year of data.

Furthermore, consider the case for a new experiment to be constructed
with 100 times the fiducial mass. What constraints is it likely to
achieve? Here we can use $\sigma \sim \sqrt{N}$ and the 1-sided
Gaussian approximation that 90\% CL corresponds to
$\sim$1.28$\sigma$. This means that after 1 year of running we would
typically expect a bound of 1.28$\sqrt{900}$/100kg = 0.38 events per
kg per year, which is still not as restrictive as the first run with a
substantially inferior experiment.

Therefore, using the value of the derived frequentist bounds alone to assess the relevance of a given experimental observation leads to a counter-intuitive behavior that does not appear to place the measurement in the desired context.
This demonstrates that individual frequentist bounds do not actually relate
to restrictions on the model and do not even necessarily represent a
measure of how sensitive or informative one experimental measurement
is relative to another one.  Frequentist bounds only take on meaning
in these regards in the actual presence of a large ensemble measurements, 
but not individually. Compare this with the Bayesian case below.

\subsection{Bayesian}

Here we will assume that all non-negative rates have an {\em a priori}
equal probability density and, accordingly, choose a uniform prior in $S$ for $S \ge 0$.
While numerical values for the derived bounds may
be modified under a different assumption, their qualitative behavior
will remain the same. For the case at hand (9 events expected, 5
observed), the uniform prior assumption yields a value of $S_{up}$ = 3.88
events per kg per year, which is $\sim$14 times larger than the
standard frequentist bound.

Now consider again what would happen if we decided to run the
experiment for another year, once more assuming that we would likely
end up with 14 events with 18 expected over two years. The above
formalism would then result in a bound of 5.89 for the 2-year run, or
a 90\% CI rate limit of 2.945 per kg per year, which is noticeably
better than before.

For the case of a 1-year exposure of an experiment with 100 times the fiducial mass, 
the limit would approach 0.38 events per kg per year (as before), which is very significantly better. 

Thus, Bayesian bounds on the actual model behave as would be expected
for something that reflects the success and relevance of a given
experimental measurement, indicating that it is generally beneficial to
run for longer and build better experiments under such scenarios.

A Bayesian calculation will also result in a more stringent upper bound for downward fluctuations because these bounds make explicit use of the constraint 
that the number of background events actually detected cannot be larger 
than the total number of observed events ({\em i.e.} the denominator of 
Equation \ref{BayesPoisson2}). The most stringent bounds therefore always 
occur when $n=0$, since the number of backgrounds is then also known 
to be identically zero for this observation. Hence, Bayesian intervals are independent of the expected background rate for such cases. However, this is not so for the frequentist case, which has no such normalization and where much larger variations for individual 
measurements are allowed because less likely measurements carry 
inherently less weight in the ensemble of other possible outcomes that 
defines the coverage. A frequentist limit based on an observed non-zero 
$n$ can, counterintuitively, even be more stringent than a limit for $n=0$, 
depending on the expected background levels for each case.

\subsection{Feldman-Cousins}

The Feldman-Cousins (F-C) bound on the scenario of 9 expected background events and 5
observed events yields a 90\% CL value of $S_{up}$ = 2.38. This is $\sim$9
times larger than the standard frequentist bound but a factor of
$\sim$1.6 smaller than the Bayesian uniform prior value, so clearly has a
different interpretation than either of these other approaches. It
neither refers to bounds on the physical model, as in Bayesian limits,
nor are the sets of observables selected to define the statistical coverage 
necessarily in direct proportion to the frequency of possible measurements, 
as in standard frequentist intervals.

This can be seen even more clearly by considering a more extreme
example in which 5 background events are expected but no events are
observed during the run. In the Bayesian approach, the background is
known to be identically zero for this one and only measurement,
leading to a 90\% CI upper bound on an average signal of 2.3 (uniform
prior case). For the standard frequentist approach, concerned with
the frequency of the observed number of counts in a large ensemble of
experiments, an empty interval is returned for a 90\% CL since this
observation has a probability of much less than 10\% even under the zero signal
hypothesis, so no positive signal strength can accommodate the
criteria. However, for F-C, the following Table \ref{fluctuation} shows the upper
limits obtained for different confidence intervals.

\vskip 0.2in

\begin{table}[H]
\begin{center}
\begin{tabular}{|l||c|c|c|c|} \hline
Confidence Level & 68.27\% & 90\% & 95\% & 99\% \\ \hline
Upper Bound on $S$ & 0.19 & 0.98 & 1.54 & 2.94 \\ \hline
\end{tabular}
\end{center}
\caption{Feldman-Cousins upper bounds on S when 5 counts are expected but none are observed (an occurrence with a statistical frequency of less than 1\% for S=0).}
\label{fluctuation}
\end{table}

It may look odd to have intervals for $S$ with up to nearly 3 signal events allowed for these
confidence levels when, for an expected background of 5 events, the
frequency with which no counts would be observed even if S were
identically zero is only 0.67\%. However, this is a reflection of how 
the acceptance region in the observable has
been distributed in a way that is not proportional to the frequency of
possible observations and that the statistical coverage
only takes on meaning for
the ensemble.

As mentioned previously, Feldman and Cousins did recognize the
problematic nature of limits such as those in Table~\ref{fluctuation}
and recommended stating both the limit and the expected sensitivity,
which in this case is 5.18 at 90\% C.L. for $S=0$, more than 5 times
larger than what appears in the table.  This large difference between
the expected sensitivity and the limit is a warning flag that the
limit should be interpreted with extreme caution.  Had one event been observed,
the Feldman-Cousins 90\% limit would be 1.22, and for $n_{obs}=2$ it
would be 1.73, all of which are noticeably lower than the expected
sensitivity. However, the probability of $n_{obs} \le 2$ is by no
means negligible (12.5\%).  The fact that the Feldman-Cousins
procedure results in limits that appear restrictive but are actually less likely to contain the true value
of the parameter in such a significant fraction of cases is ultimately unsatisfactory.

Various attempts have been made to modify the Feldman-Cousins
procedure to improve its performance for downwards fluctuations in
background.  Roe and Woodroofe presented an approach in which the
likelihood is replaced by a ``conditioned'' likelihood, where the
confidence intervals are constructed using conditional probabilities
given the constraint that the number of background events cannot
exceed the total number of observed events~\cite{rw}.  While this
generates satisfactory results in the specific case of Poisson
processes, Cousins has shown that this procedure does not generalize
well, and gives unsatisfactory results for a continuous Gaussian
variable near a physical boundary~\cite{rwcousins}.  In response, Roe
and Woodroofe have proposed using Bayesian credible intervals in a
similar way to what we will discuss in this paper, and have explored
some of their coverage properties~\cite{rw_bayes}.

\subsection{Example of Behavior Under an Improved Analysis}

As one more example to compare the behavior of upper limits for the
different approaches, first consider the case where 5 backgrounds are
expected and 2 events are observed. The resulting 90\% CL/CI upper bounds
are given in the first column of Table 2. Now assume that an improved
analysis technique is developed that is expected to reduce the
background levels by a factor of 10 while not impacting the efficiency
of signal detection. When this is applied to the same data set, the
events previously observed are cut. The new 90\% CL/CI upper bounds are
then re-computed and given in the second column of Table
\ref{improved_analysis}.


\begin{table}
\begin{center}
\begin{tabular}{|l||c|c|} \hline
                                  &                & Improved Cuts: \\ 
                                  & B=5, n=2 & B=0.5, n=0 \\ \hline
Standard Frequentist & 0.32 & 1.8 \\ \hline
Feldman-Cousins & 1.73 & 1.94 \\ \hline
Bayesian & 3.13 & 2.3 \\ \hline
\end{tabular}
\end{center}
\caption{90\% CL/CI upper bounds on S when 5 background counts are expected and 2 events are observed, compared with those for an analysis with 10 times better background rejection.}
\label{improved_analysis}
\end{table}

\enlargethispage{\baselineskip}

For both standard and F-C frequentist approaches, paradoxically, the
improved analysis actually results in a worse constraint on S. Only
the Bayesian limit improves, behaving more as would be intuitively expected 
under this scenario. This example goes to further
illustrate the point already made: that individual frequentist bounds
neither relate to restrictions on the model, nor do they necessarily
represent a measure of how sensitive or informative one experimental
measurement is relative to another one.

Pragmatically, we believe that any useful method for producing limits
must do so for {\em any} data observation, with a meaning that is easy to
interpret and provides a reasonable, robust and intuitive basis on
which to compare results.  Both the standard and Feldman-Cousins
methods appear to fail in this regard a non-negligible fraction of the
time, and the suggestion that this problem can be dealt with by simply
quoting the expected sensitivity and leaving the interpretation to the
individual's judgement does not seem like a viable way to proceed.

\section{Example of Issues with Intervals in the presence of a clear signal}

Up to now, examples have focused on potential issues of misinterpretation related to upper bounds. We give here an example where a 2-sided interval construction for a clear signal can also lead to complications for a frequentist approach.

Consider the case of an ultra-high energy neutrino detector, such as
IceCube \cite{IceCube}, looking for signs of extra-terrestrial
neutrinos from astrophysical sources. Assume that the instrument has a
known Gaussian energy resolution and that one event with an apparent
energy well beyond expectations for atmospheric neutrinos is observed.
Say we now wish to construct a confidence interval for the energy of
the neutrino itself (as opposed to the deposited energy). However, the
likely energy of the event is strongly dependent on the index of the
underlying energy spectrum, which is unknown. For example, the
observation is much more likely to have resulted from the fluctuation
of a lower energy event if the underlying differential neutrino
spectrum was proportional to $E^{-3}$ as opposed to $E^{-2}$ or
$E^{-1}$. Without knowing this index, it is therefore not possible to
uniquely define a hypothetical ensemble of repeated measurements, and
frequentist bounds on the deposited energy alone can be misleading if
incorrectly interpreted in terms of the neutrino energy. One could
treat the index as a nuisance
parameter but, as shown later, the associated uncertainties still cannot be propagated in a self-consistent manner using a purely frequentist framework. However, Bayesian bounds are well-defined, where the dependence on the assumed spectral prior is made explicit and the sensitivity to this choice can be shown.

While this example was chosen as a particularly clear case, all intervals are subject to this issue at some level. If the choice of prior is obvious or does not matter, Bayesian bounds are unambiguous. If the choice of prior is not clear and leads to differences in interpretation, a Bayesian construction is explicit regarding this context, whereas the misinterpretation of a frequentist interval as bounds on a model can lead to erroneous conclusions that are effectively based on an assumed but hidden prior.

\section{Comparison of Experimental Limits}

It has already been noted that frequentist intervals do not necessarily reflect the relevance of individual experimental measurements. However, this issue is worth further discussion since the comparison of derived frequentist intervals from different experiments are often made on exclusion plots etc., which can lead to erroneous conclusions. 

\begin{figure}
\includegraphics[width=85mm]{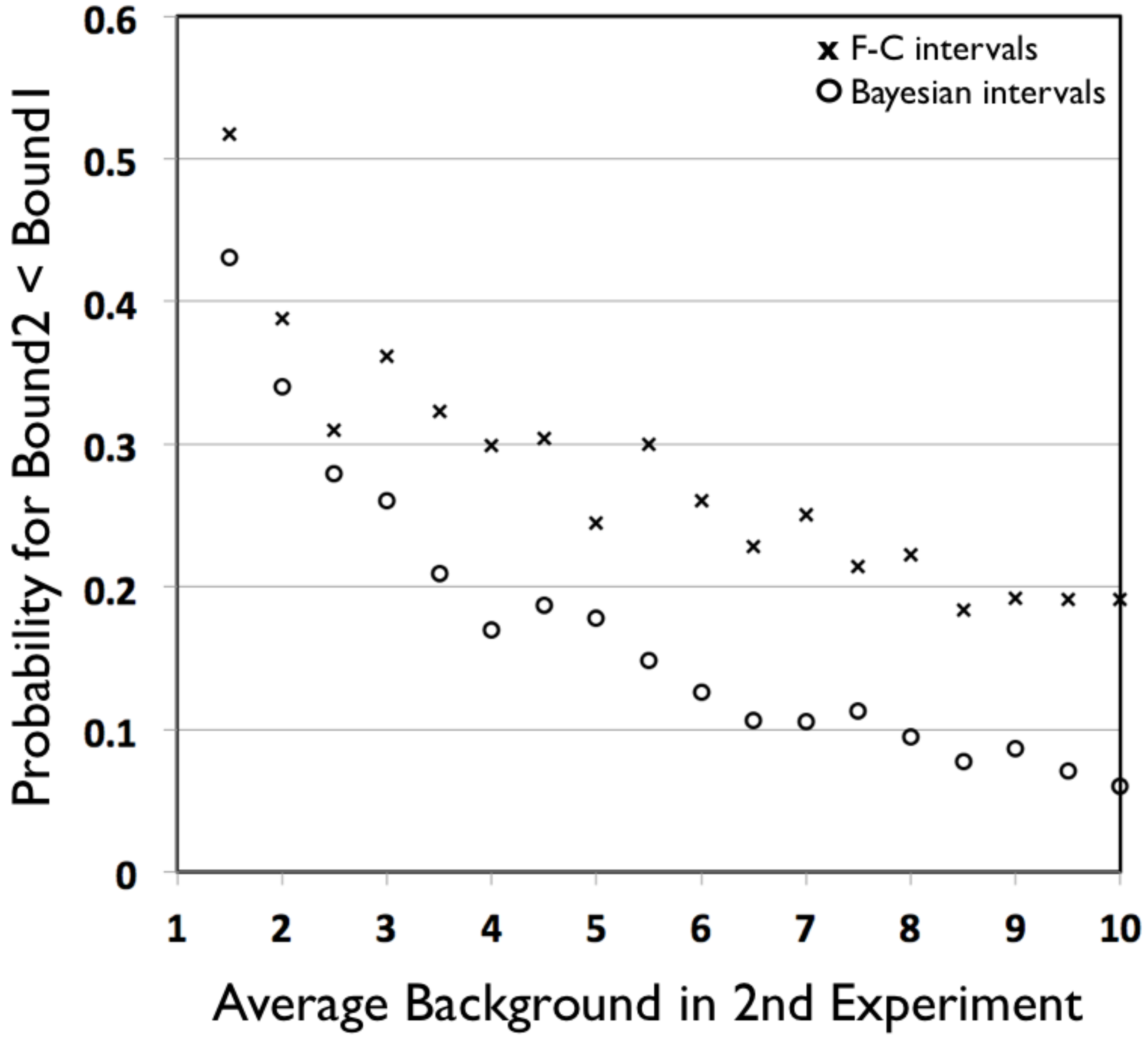}
\caption{The fraction of times that the derived 90\% CL/CI upper
  interval bounds for Experiment \#2 (higher background) are found to
  be more restrictive than those of Experiment \#1 (background of 1
  event) as a function of the average background level for Experiment
  \#2. Cases for uniform-prior Bayesian (circles) and Feldman-Cousins
  (crosses) constructions are shown. Some unevenness due to Poisson
  quantization 
is visible. }
\label{better_bounds}
\end{figure}

To illustrate this, consider the scenario of two counting experiments making observations to place bounds on an possible signal. The first of these has an expected background of just 1 count, while the second suffers from a higher average background level. 
Now consider the ensemble of comparisons between 90\% CL/CI upper interval bounds
for Experiment \#1 and Experiment \#2 under the zero signal
hypothesis. Figure \ref{better_bounds} plots the fraction of times the
derived upper bound for the interval of Experiment \#2 is found to be
smaller than that of Experiment \#1 as a function of the
average background level for Experiment \#2. Some unevenness due to
Poisson quantization 
can be seen, especially for lower background
numbers. However, in all cases, this fraction is substantially greater
for Feldman-Cousins (crosses) as compared with Bayesian (circles)
intervals, with the difference as large as a factor of $\sim$3 at
higher background levels. In other words, bounds derived from
Feldman-Cousins are less likely to reflect the relative sensitivities
of experiments.

\begin{figure}
\includegraphics[width=80mm]{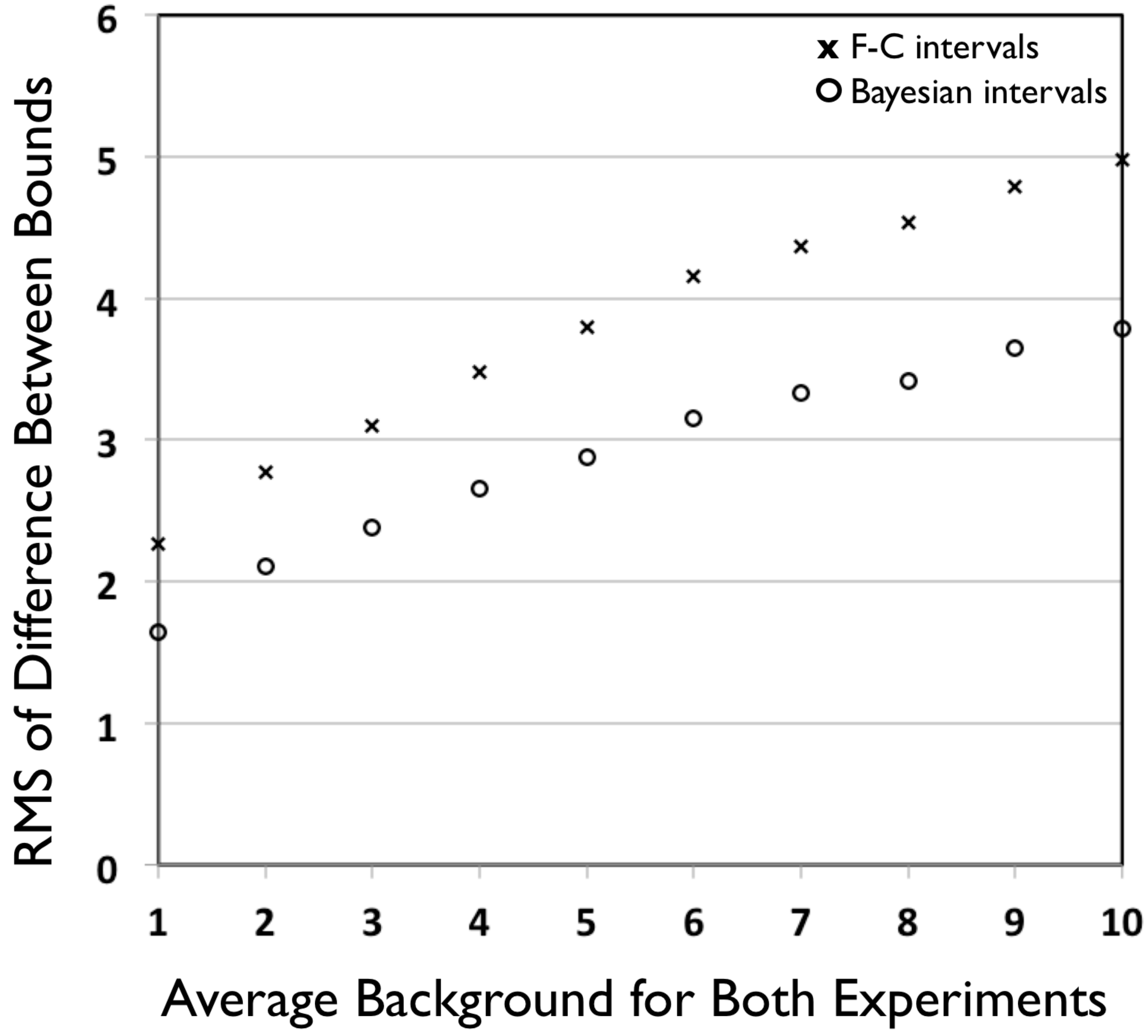}
\caption{The RMS deviation between 90\% CL/CI upper interval bounds from two experiments with the same expected background level. Cases for uniform-prior Bayesian (circles) and Feldman-Cousins (crosses) constructions are shown for the zero signal hypothesis as a function of expected background level. Notably larger fluctuations in the derived bounds are seen for the Feldman-Cousins case.
}
\label{RMS}
\end{figure}

Another indicator of the robustness of derived bounds under the 
hypothesis of zero signal is the size of the RMS deviation associated with the
difference between two such bounds, obtained for pairs of experiments
with identical expected background levels. This is shown in Figure \ref{RMS} 
for the 90\% CL/CI upper bound in intervals for uniform-prior Bayesian (circles) and Feldman-Cousins (crosses) constructions as a function of the average background level for both experiments. In all cases, the Feldman-Cousins bounds correspond to RMS values that are over 30\% larger than for the Bayesian case, indicating that these are noticeably less robust, being more likely to yield apparent discrepancies between different experiments and to change under repeated observations.

These issues are more than of esoteric interest, and several examples can be found where the representation of experimental results appear to run into difficulties when couched in a frequentist context.

\subsection{KARMEN II}

The Karmen II neutrino oscillation experiment observed zero events during its 
initial run between February 1997 and April 1998, where the expected background was 2.88$\pm$0.13~\cite{karmen}. Derived Feldman-Cousins bounds consequently yielded an upper limit of 1.07 events at the 90\% CL, more than a factor of two more restrictive than uniform-prior Bayesian bounds would have produced. This led to numerous incorrect statements in the literature concerning constraints on model parameters and the widespread promulgation of exclusion plots comparing frequentist bounds that erroneously suggested significantly better constraints relative to other experiments than were justified. Further data gathered up to March 2000 quadrupled the statistics and, after fluctuations took their due course, 11 events were observed in the full data set compared to 12.3$\pm$0.6 expected~\cite{karmen2}. This produced nearly identical constraints to the previous set, which had only 1/4 the exposure. Had uniform-prior Bayesian bounds been used instead, the experimenters would have found that their constraints were initially less restrictive, but then improved by a factor of $\sim$2 when the statistics were quadrupled, in line with expectation.

\subsection{LEP and LHC (The CL$_{s+b}$ Method)}
Members of the ALEPH, DELPHI, L3 and OPAL collaborations recognized the potential difficulties in interpreting frequentist limits in the face of fluctuations. In their joint 2003 paper `Search for the standard model Higgs boson at LEP,' \cite{LEP}, the authors note the following regarding frequentist intervals: ``{\em ...this procedure may lead to the undesired possibility that a large downward fluctuation of the background would allow hypotheses to be excluded for which the experiment has no sensitivity due to the small expected signal rate.}'' 
Their solution was to re-define their ``frequentist'' upper bounds based on the ratio of the chance probability for the observation under a given signal+background hypothesis to that under the hypothesis of background alone \cite{Read}. In the context of a simple counting experiment, this ``CL$_{s+b}$ method'' would therefore take the form of Equation \ref{BayesPoisson2} which, in fact, is a Bayesian bound with uniform prior. The equivalence with a Bayesian bound also holds for a Gaussian with a prior uniform in the mean, and can sometimes be found for other distributions with different choices of prior. However, this equivalence is not true in general and, in particular, does not hold for the likelihood ratio test statistic often used to search for new particles. 

While the CL$_{s+b}$ statistic avoids bounds that may appear overly strict for negative fluctuations, the interpretation of the associated confidence levels is unclear and, in fact, is variable, depending on the test statistic. It does not guarantee frequentist coverage, nor does it necessarily provide well-defined bounds on model parameter values themselves (and even in cases where there is a Bayesian equivalent, the form of the prior is not explicitly evident). Nevertheless, the technique is now ubiquitous amongst LHC experiments as well, being used (as with LEP) essentially as a binary assessment as to whether an observation is significant. If so, a 2-sided Feldman-Cousins interval is then typically quoted, though this scheme now violates the principles on which the Feldman-Cousins method is predicated ({\em i.e.} that the nature of the interval is automatically determined by the construction). For such a binary assessment, it is unclear what advantage this provides over a simple p-value (test of consistency with the zero signal hypothesis). Beyond this, if one wished to constrain model parameter values themselves with well-defined confidence levels, appropriate Bayesian bounds could instead be derived.

It is worth emphasizing again that the issue of negative fluctuations is entirely a consequence of misinterpreting frequentist bounds or, equivalently, using the wrong construction to answer a Bayesian question.  We find it curious that, in order to apparently avoid a philosophical issue about providing an unambiguous interpretation, the authors have instead opted to use a scheme without any consistent interpretation at all. A more straightforward approach would be to confront the specific nature of the questions being posed and then adopt an appropriate, well-defined and consistent mathematical formalism.

\subsection{ZEPLIN III}

In 2009, the ZEPLIN III collaboration published their first bounds on dark matter~\cite{zeplin}. The limits were largely based on a large region within the signal box where no events were observed. The authors noted that the fit expectation for the average background level was greater than this and, together with the difficulty in quantifying some systematics associated with the background extrapolation, this ``compromised'' the use of frequentist techniques, such as maximum likelihood or Feldman-Cousins. Consequently, they instead used the maximum value allowed for a Feldman-Cousins 90\% CL interval of 2.44 (the value at $n=B=0$).  Clearly, while this may be seen as a ``conservative estimate'' of a frequentist bound that must be greater than 90\%, the meaning of the confidence level beyond this is simply not defined.
Had the authors instead used a Bayesian approach with a uniform prior,
they would have arrived at a bound of 2.3 at 90\% CI  --- a value that is, in fact, well-defined for this case.

\subsection{EXO-200}

The EXO collaboration published first results from the EXO-200 neutrinoless double beta-decay experiment in 
2012~\cite{exo}. In the $\pm$1$\sigma$ energy resolution window around the endpoint, 1 event was observed where a background of 4.1$\pm$0.3 counts was expected. Using a spectrum fit, the authors derived a bound of $<$2.8 total signal counts at the 90\% CL, corresponding to a lower bound to the half-life for 0$\nu\beta\beta$ of $1.6\times10^{25}$ years. 
A Feldman-Cousins bound based on the  $\pm$1$\sigma$ bin would have
yielded a limit of less than 2.0 signal counts at 90\% CL (accounting
for the 68\% signal efficiency of the bin), corresponding to an even
more restrictive 90\% CL lower bound for the half-life due to the
negative fluctuation of $2.2\times10^{25}$ years. On the other hand, a
Bayesian bound with a prior uniform in counting rate based on the
$\pm$1$\sigma$ bin would have yielded a limit of less than 4.0 signal
counts, corresponding to a 90\% CI lower bound to the half-life of
only $1.1\times10^{25}$ years 
--- 
seemingly less restrictive than the Feldman-Cousins bound by a factor of two.

The EXO data falls exactly into the category of paradoxical situations for frequentist intervals previously described, where improved background rejection and/or longer periods of data collection would likely result in less restrictive bounds than for the initial case. And, in fact, an update of EXO-200 results published in 2014 using two years of data with quadruple the exposure of the initial result appeared to actually suggest a positive $\sim 1.2\sigma$ fluctuation in the larger data set that accentuates this effect \cite{exo2014}. Within the $\pm$1$\sigma$ energy resolution window around the endpoint, 21 events were observed where a background of 16$\pm$2 counts was expected \cite{exo2}.
All approaches derive similar bounds for this case: 
the authors
derived a 1-sided 90\% CL lower bound to the 0$\nu\beta\beta$
half-life of $t_{1/2}>1.1\times10^{25}$ years by applying Wilks'
Theorem to a likelihood analysis, which is a factor of 1.45 less
restrictive than the initial result. A Feldman-Cousins analysis based
on the $\pm$1$\sigma$ bin would yield a bound of
$t_{1/2}>1.3\times10^{25}$ years, a factor of 1.7 less restrictive
than the F-C bound from the initial result. However, a Bayesian bound,
with a prior uniform in counting rate and based on the  
same bin yields a value of $t_{1/2}>1.4\times10^{25}$ years or, using the appropriate integration of the posterior probability derived from the provided likelihood curve assuming a uniform prior,  a value of $t_{1/2}>1.2\times10^{25}$ years. Both Bayesian calculations are modestly {\em more} restrictive than the initial Bayesian result, thus better reflecting the relevance of the measurements and providing a substantially more stable basis for comparison in the face of these background fluctuations. These results are summarised in Table \ref{exo}.

\begin{table}
\begin{center}
\begin{tabular}{|l||c|c|c|} \hline
                           & Spectrum  & F-C  & Bayesian \\
                           & + Wilks'         &    ($\pm1\sigma$ bin)    &    ($\pm1\sigma$ bin)    \\ \hline
 Data Set 1         &  $>$1.6 &  $>$2.2 & $>$1.1 \\ \hline
 Data Set 2         &  $>$1.1 &  $>$1.3 & $>$1.4 \\
                           &              &              & (integration: $>$1.2) \\ \hline
 Set 1 / Set 2      &   1.45 & 1.7 & 0.78 (0.92) \\ \hline
\end{tabular}
\end{center}
\caption{Derived 90\% CL/CI lower bounds on the half-lives for 0$\nu\beta\beta$ in units of $10^{25}$ years from EXO data using various approaches. `Data Set 1' is from the initial 2012 publication \cite{exo} and `Data Set 2' is from the 2014 publication with quadruple the exposure \cite{exo2014}. The ratios of bounds between data sets are also given. Notably less restrictive frequentist bounds result from the larger data set owing to background fluctuations, whereas the Bayesian bound modestly improves.}
\label{exo}
\end{table}

\vskip 0.4in

\enlargethispage{\baselineskip}

These are just a few obvious examples of cases sampled across a number of different areas in particle physics. However, the fact is that {\em all} such comparisons of experimental results using frequentist bounds are sensitive to these issues at some level.

\section{Issues Associated with the Treatment of Nuisance Parameters}
\label{nuisance}

The concept of frequentist coverage presents particular challenges
when trying to incorporate the effects of other unknown parameters.  A
confidence interval construction is said to have, say, 90\% coverage
if, for any true value of a parameter $\theta$ that is to be
estimated, an ensemble of repeated experiments would result in
constructions that would contain this value in 90\% of the repetitions.

Consider now the case that the likelihood $L(x|\theta,\xi)$ depends on
a second parameter $\xi$.  Here $\xi$ may either be a parameter of
physics interest, or a nuisance parameter representing the effects of
a systematic uncertainty.  A common experimental problem is to
determine a 1D confidence interval for $\theta$, independent of the
value of $\xi$.

Standard frequentist techniques can readily define a 2D confidence
region in the $\theta,\xi$ plane so that, for any point
($\theta,\xi$), the generated region will contain that point in 90\%
of random trials. But these techniques do not provide any satisfactory way of
producing a 1D confidence interval for $\theta$, independent of $\xi$,
with a desired level of coverage.

Two possible definitions of coverage may be considered in this
case. The ``strong'' definition of coverage would be that the 1D
interval generated by the method should contain the true value of
$\theta$ in 90\% of cases, for any values of $\theta$ and $\xi$.  This
would be the desired definition of coverage for a purely frequentist
construction, since $\xi$ may represent a parameter whose value is
constant but unknown, and not subject to fluctuations from trial to
trial.

One may also consider a ``weak'' coverage requirement.  In this
approach, $\xi$ is thought of as a random variable that may have a
different value every time the experiment is done (although this is
not always the case in actuality). If the
frequency distribution for $\xi$ is known and denoted by $f(\xi)$,
then we could less stringently require that the frequency for the 1D
interval generated by a random measurement $x$ to contain the true
value of $\theta$ is 90\%, {\em averaging over $\xi$}.  For some fixed true value of
$\xi$, the 1D confidence interval generated might not have the desired
90\% coverage, but since the true value of $\xi$ is, by assumption,
not known, we are content if:
\begin{equation}
 \int d\xi f(\xi) \alpha_\theta(\xi) = 0.9 
\label{eq:weak}
\end{equation}
where $\alpha_\theta(\xi)$ is the coverage at true value $\theta$ for
a particular true value of $\xi$:

\begin{equation}
\alpha_\theta(\xi) = \int_{x \in R} dx P(x|\theta,\xi).  
\end{equation}

Here $R$ denotes the region in the measurement space $X$ determined by
whatever ordering principle is used to construct the confidence
intervals.

If $\alpha_\theta(\xi)$ is a constant, not depending on $\xi$, then
the coverage is in fact independent of the nuisance parameter and the
strong definition of coverage is obtained.

Note that there is no essential difference between the case where
$\xi$ is a true nuisance parameter versus the case that we wish to
incorporate the effect of one ``physics'' parameter in the projected confidence region for another, such as generating a 1D
confidence interval for the neutrino mixing parameter $\theta_{23}$
from a likelihood function that depends on $\theta_{23}$ and $\Delta
m^2_{32}$.

\subsection{The Frequentist Minimization Procedure} 

The commonly recommended frequentist prescription for eliminating a
nuisance parameter is the ``profile'' method, in which the
profiled likelihood is generated by maximizing $L(x|\theta,\xi)$
over $\xi$ for each fixed value of $\theta$:

\begin{equation}
L(x|\theta) = \max_\xi L(x|\theta,\xi)  
\end{equation}

This suggests that, for example, to generate a Feldman-Cousins confidence interval in the presence of a nuisance parameter, we should form the following likelihood ratio:

\begin{equation}
\Lambda_\theta (x) = \frac{L(x | \theta, \hat{\xi}_\theta (x))}{L(x|\hat{\theta},\hat{\xi})}
\label{eq:lambda}
\end{equation}
Here in the numerator, $\hat{\xi}_\theta (x)$ is the value of $\xi$ that maximizes the likelihood for a fixed value of $\theta$.  In the denominator, $\hat{\theta}$ and $\hat{\xi}$ are the values of these parameters that globally maximize the likelihood.  In all cases 
$\Lambda \le 1$.

For any value of $\theta$, there is a critical value $c_\theta$ for which $\theta$ will be included in the confidence
interval if $\Lambda_\theta(x) > c_\theta$.  The value of $c_\theta$ is chosen by construction so that the frequency with which this selection occurs is 90\%.

Ideally, we would want the critical value $c_\theta$ to be independent of $\xi$.  In that case, the strong coverage condition holds, and the integrity of the profiled 1D frequentist interval is maintained.

\subsection{A Simple Example}

Suppose we perform a single measurement of a quantity expected to
follow a Gaussian distribution with mean $\theta + \xi$ and an RMS of
1, and that the result of this measurement is $x$.  In this case,
$\theta$ and $\xi$ are degenerate.  Let us therefore suppose that we
have further knowledge that $\xi = A \pm 1$, with $\xi$ following a
Gaussian distribution with mean $A$ and RMS 1.  
This could result from a previous measurement
of $\xi$, such as from a calibration run.  The (unnormalized) joint
likelihood function is therefore:
\begin{equation}
L(x|\theta,\xi) = \exp[-\frac{1}{2}(x - (\theta+\xi))^2]  
\exp[-\frac{1}{2}(\xi - A)^2]  
\end{equation}

It is trivial to see that this is globally maximized for $\hat{\xi} =
A, \hat{\theta} = x-A$, at which point 
$L(x|\hat{\theta},\hat{\xi})=1$.

For any fixed value of $\theta$, the likelihood is maximized at
\begin{equation}
\hat{\xi}_\theta(x) = \frac{x -\theta + A}{2}
\end{equation}
This is just the arithmetic average of the value $\xi = x-\theta$ that maximizes the first factor in the likelihood, and the best fit value $\xi = A$ from the second factor.  If the two Gaussian terms in the likelihood had different $\sigma$'s, this would instead be a weighted average.

Inserting this expression into Eq.~\ref{eq:lambda} gives 

\begin{equation}
\Lambda_\theta(x) = \exp [-\frac{1}{4}(x - A - \theta)^2].
\end{equation}

This is the ordering parameter.  For any fixed ($\theta,\xi$), we can predict the distribution of $x$ and, hence, of $\Lambda_\theta(x)$, and determine the critical value $c_\theta$ for $\Lambda_\theta(x)$ such that:
\begin{equation}
\frac{ \int dx L(x|\theta,\xi) H(\Lambda_\theta(x) - c_\theta)} { \int dx L(x|\theta,\xi)} = 0.9
\label{eq:coverage}
\end{equation}
where $H$ is the Heaviside step function to insure that we include $\theta$ in the confidence interval only if $\Lambda_\theta(x) > c_\theta$.

Since the distribution of $x$ depends on $\xi$ but $\Lambda_\theta(x)$ does not, it is clear that Equation \ref{eq:coverage} cannot be guaranteed to hold for all $\xi$. Therefore, the confidence interval construction does not satisfy the strong coverage condition---the procedure does not yield confidence intervals that give the correct coverage for all combinations of $\theta$ and $\xi$. 

In this case, one might at least try to err on the side of caution by choosing the smallest critical value that is obtained for any $\xi$, giving the desired coverage level for that one value of $\xi$ and giving over-coverage for other values.   (For example, Cranmer has
proposed including in the 1D interval any value of $\theta$ for which
the standard 2D confidence region is not empty for at least one value
of $\xi$~\cite{cranmer}.) However, this method not only is extremely likely to give over-coverage for almost all true values of $\xi$, but also may give overly large intervals dictated by the most extreme possible value of $\xi$.

The alternative is to give up on the strong coverage condition and settle for ``weak coverage'', as per Equation~\ref{eq:weak}. To achieve this, one can simply interpret the likelihood function as a joint probability distribution for $x$ and $\xi$.  Drawing values for $x$ and $\xi$ randomly from this distribution (for fixed $\theta$), one can, for example, then calculate the distribution for $\Lambda_\theta(x)$ by Monte Carlo and determine the appropriate critical value to give the desired coverage.  Correct coverage of the ``weak'' kind is then obtained with the following meaning: if the experiment were done a large number of times, and if $\xi$ had a different random value for each trial with a joint probability distribution $L(\xi,x|\theta)$, then 90\% of the generated intervals would contain the true value of $\theta$.

This is clearly a hybrid approach.  In order to create a
``frequentist'' interval with desired coverage for $\theta$, we are
integrating out the nuisance parameter $\xi$.  In doing this, we are
forced to treat $\xi$ in a Bayesian way, with an assumed prior
distribution, and must partly abandon the frequentist paradigm.  This
is of course exactly the approach of
Cousins-Highland~\cite{cousins-highland}, and its performance and that
of profiling the likelihood have been explored by a number of authors
\cite{conrad,rl}.   
The contribution of this Bayesian aspect to the
overall confidence interval is not necessarily small since, for
example, it is a common goal to run experiments to the point where
systematic uncertainties dominate.

Even if one still chose to accept a pseudo-Bayesian way of getting rid of nuisance parameters in order to create pseudo-frequentist intervals on the remaining parameters of interest, the incursion of Bayesian philosophy cannot necessarily be so simply contained. Consider the case of a neutrino oscillation experiment with no systematic uncertainties, and sensitivity to two oscillation parameters $\Delta m^2$, $\theta$.  While one can readily produce frequentist confidence regions in the 2D contour plane, what happens if we want to quote a 1D limit on either of the parameters?  This is mathematically identical to eliminating a nuisance parameter, although, in this case, the parameter is actually 
one of physical interest.  Except in special cases, it is not possible to produce 1D frequentist confidence intervals with correct
coverage for all values of the other parameter.  The best that can be hoped for is to achieve ``weak'' coverage, but this then implies marginalising in a Bayesian way over the other parameter (as, for example, in \cite{t2k}).  One is forced to be Bayesian about any parameter that is being eliminated from the problem, or else abandon the notion of defining a consistent statistical coverage.

We suggest that, rather than seeking such work-arounds and compromises to the frequentist treatment of nuisance parameters, it is prudent to instead ask why such steps are necessary in the first place. If the mathematical framework is inconsistent, this may suggest that the thinking leading up to the use of such an approach is also inconsistent and, therefore, likely to lead to further difficulties and misinterpretations.

\subsection{Bayesian Treatment of Nuisance Parameters}

Nuisance parameters present no special difficulties for a Bayesian
analysis.  If
$L(D|\theta,\xi)$ is a likelihood function for a datum $D$ depending
on a nuisance parameter $\xi$, any constraints on this parameter
are easily included as part of the prior in Bayes' Theorem.  For
example, $\xi$ might represent the rate of a background process,
perhaps measured in a separate calibration or side channel.  A
probability distribution $g(\xi)$ can then be assigned for $\xi$
which, together with a prior $f(\theta)$ for the physics parameter
$\theta$, can be used as priors in Bayes Theorem to give a joint
posterior distribution for both $\theta$ and $\xi$:
\begin{equation}
P(\theta,\xi | D) \propto L(D|\theta,\xi) f(\theta) g(\xi)  
\label{eq:bayessys}
\end{equation}
Dependence on the unwanted parameter $\xi$ can then be removed simply
by integrating the posterior distribution with respect to $\xi$.

\vskip 0.2in

It is worth noting that because the likelihood function
$L(D|\theta,\xi)$  depends on $\xi$, the measurement itself may
often contain useful information on the true value of $\xi$.
Incorporating prior information on $\xi$ along with information
derived from $D$ through the likelihood makes full use of all of the
information about $\xi$ that is available.  This is often a superior
approach to Monte Carlo methods in which random values of $\xi$ are
drawn from the distribution $g(\xi)$, and then used to fit for
$\theta$, with $\xi$ held constant in each fit.  If the Monte Carlo approach is used,
the results of fitting for $\theta$ using each random throw of $\xi$
should ideally be weighted by the calculated likelihood for that value of $\xi$.

Choosing appropriate priors for nuisance parameters representing
systematic uncertainties is also generally straightforward in practice.  If the
constraint on the nuisance parameter is the result of an
independent measurement, the likelihood function for
that measurement can be an appropriate choice of prior for
$g(\xi)$ in Equation~\ref{eq:bayessys} (possibly further modified by
any theoretical priors on $\xi$ itself).  Physical boundaries, such as
requiring background rates to be non-negative, are easily incorporated
by setting the prior to zero in the unphysical region and, in fact,
should always be included to prevent unphysical behavior in the
posterior distribution.  In the case that there is no previous
measurement upon which to base a prior for the nuisance parameter, the
guidelines in Section 
\ref{sec:priors} (``Choice of Bayesian Priors'')
may be used.

\ref{marginalize} contains further discussion of
the mathematical relationship between integrating over a nuisance
parameter vs. maximizing the likelihood with respect to one.

\section{Issues with Unification}
\label{unification}

The paper of Feldman and Cousins advocates the use of a ``unified
approach," in which the formalism of the interval construction itself
dictates whether upper bounds or two-sided confidence intervals are
given. The argument for this is to avoid the problem of
``flip-flopping," whereby different experimenters choose for
themselves when to quote a given type of interval based on the result,
leading to a small statistical bias in frequentist coverage in some
cases if one were to do an unfiltered survey of only those frequentist
intervals reported. It should be emphasized again that this is a
purely frequentist issue 
--- 
Bayesian intervals are immune to such
effects as they are not defined with respect to ensembles.

The frequency with which signals are excluded can, for borderline
cases, be as much as 50\% higher than would be inferred from the
nominal confidence level 
(see Appendix B). 
In other words,
an ensemble of 90\% CL bounds may only have 85\% coverage. 
This deviation from nominal coverage is of similar magnitude to the
inherent variations in frequentist coverage due to quantized Poisson
statistics (see Appendix A).
In practice, it is also the case that
any such biases are often dwarfed by other factors, including
difficulties in assessing and propagating systematic uncertainties,
accounting for look-elsewhere effects, and various details of the
particular analysis approach employed. In addition, it should be
considered that results are generally not taken purely at face value,
but are often re-analyzed and combined with other results when
appropriate and that the effect in question diminishes as significance
levels move away from the cross-over region (where measurements are
generally viewed conservatively in any case, independent of what type
of interval may be quoted). Therefore, the potential impact of
``flip-flopping" is, in fact, not particularly significant in relative
terms.

On the other hand, the adoption of a unified approach imposes substantial constraints that can lead to non-trivial difficulties:
\begin{itemize}

\item The approach conflicts with the desired and scientifically well-motivated convention to quote a 90\% or 95\% CL for upper/lower bounds for results that are consistent with the zero signal hypothesis, but to only claim a 2-sided discovery interval when the zero signal hypothesis is rejected at a considerably higher confidence level (typically in excess of 3 or more standard deviations). Indeed, even in cases where a unified 2-sided interval may be shown, it is often accompanied by the phrase, ``we regard this as an upper limit'' when the significance is not judged to have passed a critical level, despite the fact that the coverage is not appropriate for such a limit. 

\item A unified approach cannot easily cope with look-elsewhere effects ({\em i.e.} trials factors). For example, if a search for gamma-ray emission from 1000 different astrophysical sources results in no event excess exceeding 3 standard deviations above the background levels, the data may be judged to be consistent with statistical fluctuations and the most appropriate things to quote are upper bounds on the possible emission from each source. However, a unified approach would instead necessitate a 3$\sigma$ detection interval for observations consistent with chance fluctuations.

\item Even in the case of a clear detection, it may still be relevant to also quote upper and lower bounds in the context of certain models. For example, some classes of models may simply place bounds on the allowed maximum luminosity of a given source. Thus, different interval constructions can be simultaneously valid and relevant for the same results, as they simply address different questions.

\end{itemize}

These difficulties appear to substantially outweigh any benefit of making what is, in the end, a minor correction to frequentist coverage. Therefore, on balance, we believe it is pragmatically advantageous to allow the nature of interval constructions to be determined by the experimenters themselves based on an assessment of scientific relevance, rather than having these dictated by an inflexible and, ultimately, inappropriate formalism.

\section{Convergence, Divergence and Confusion}

The likelihood function is central to all the approaches described and increasingly constrains the model parameters and dominates the definition of these intervals with increasing numbers of events. Hence, in the limit of large numbers of events, the dependence on the particular choice of prior becomes insignificant in the Bayesian construction, and the Feldman-Cousins ordering parameter has little effect away from a physical boundary such as the origin. All methods therefore converge on the same bounds in this region. This helps to address the question of what constitutes a sufficient ensemble of measurements in the frequentist approach for the constructed intervals to begin to reliably constrain a model. Namely, this occurs when there is a sufficient sampling to well-characterize the likelihood space of measurements, at which point such an approach would produce the same answer as for the Bayesian method. 

The deviation between approaches in the region of low numbers of events is
therefore largely a reflection of the fact that there is not yet
enough information to make a more definitive statement about the model
without supplying at least some additional constraints. The
frequentist approach in this case is to simply place the measurement
in context for some hopeful ensemble of other experiments without
trying to identify the model, while the Bayesian approach is typically
to seek out some minimal set of ``reasonable and conservative''
constraints in order to infer which model parameter values are the
most likely. These philosophies are not mutually exclusive 
---
both goals are valuable in the case of limited information and simply need to be appropriately defined and distinguished.

However, in many cases, a lack of clarity in this regard and a reticence to provide both types of information has led to confusion. An informal survey of physics journal articles suggests that, in a large fraction of cases, quoted frequentist intervals are often used to make statements regarding constraints on model parameter values, either by the authors themselves or by others in subsequent articles, even when the natures of the intervals are explicitly stated. Parameter exclusion plots, as the name implies, are instinctively interpreted as defining allowed and disallowed regions of model parameter space based on the data. This is an inherently Bayesian interpretation, yet the nature of the derived contours is not always consistent with this. We believe there is need of a more pragmatic approach which recognizes that, while it is critical to objectively convey the information content of the data, there is also a strong desire to derive bounds on model parameter values and a natural instinct to interpret things this way.

\vskip 1.0in

\section{Towards a More Relevant and Transparent Approach}

\subsection{Relevant Statements for Scientific Papers}

We start with an attempt to distill the basic statements that are desirable to make regarding the nature of results from an experimental measurement. There are typically four issues of relevance:

\begin{enumerate}

\item To present the measured value of a direct observable and an assessment of systematic uncertainties that could bias the measurement. This results in a simple, objective statement concerning the observation.

\item To address the question,``How often would a measurement `like
mine' occur under the zero signal hypothesis?'' This is a frequency question
probing statistical consistency and focusing on an observable in the
context of a single, fixed model ({\em i.e.} a Fisher-type test). For
this, the normalized PDF for observations under the zero signal hypothesis
can be appropriately integrated (including integration over any
nuisance parameters) to arrive at an assessment, {\em i.e.} a
``p-value''. There is, however, some ambiguity in what is meant by
`like mine.' For example, ``How likely is it to measure a rate this
high?" or ``How likely is it to measure a rate this far away from the
predicted value under the zero signal hypothesis?" As there is not a general
form for this, it needs to be defined in a relevant way on a
case-by-case basis. Note that this is not equivalent to defining a
frequentist interval 
--- 
the result is just a single number representing
the statistical chance of measuring a value for the observable in a
`similar range' under the zero signal hypothesis. 
Inferences based on p-values
should be treated with caution (see, for example, \cite{caldwell} for
further discussion).

\item To address the question,``What constraints do my measurements of direct observables place on model parameter values?" This is explicitly a Bayesian question and, thus, requires the application of the appropriate formalism, including the use of a prior to define the relative context of models. 

\item To objectively convey the relevant information content of the data so as to allow the impact of alternative assumptions to be evaluated, facilitate the testing of different models, and permit information from this measurement to be effectively combined with that from other experiments. Frequentist intervals are blunt instruments for this purpose that only provide a crude simplification of likelihood information viewed through a particular, non-unique filter that may be prone to misinterpretation. In practice, such intervals are rarely used in the ensemble tests for which they are relevant, being disfavored relative to combined analyses that either use the raw data or likelihood maps from different experiments. A better approach would therefore be to actually provide, to the best extent possible, the likelihood information directly. 

\end{enumerate}

The means by which to address the first two of these issues is relatively straightforward and largely uncontroversial. We will therefore now concentrate on approaches relevant for the latter two issues.

\subsection{Choice of Bayesian Priors}
\label{sec:priors}

As indicated previously, in order to use measurements to bound model parameter values, a context for these values must be provided in the form of a prior probability. This conveniently permits known, physical constraints to be imposed ({\em e.g.} energies and masses must be greater than zero; the position of observed events must be inside the detector, etc.) and allows known attributes of the physical system to be taken into account ({\em e.g.}  energies are being sampled from a particular spectrum; the relative probabilities for different event classes are drawn from a given distribution, etc.). The choice of priors in such contexts is often non-controversial. Less straightforward is the case of defining a prior within the physical region when there is no {\em a priori} knowledge of the probability distribution for a given model parameter: the so-called ``non-informative'' prior. It may seem odd to need to choose a prior at all under such circumstances, but the fact is that ``no knowledge'' is a fuzzy concept whose meaning needs to be defined. 

We should note (as others have) that it is rarely the case that there is really ``no prior information" at all, as we will generally have some knowledge of previous observations related to the current measurement. At the same time, it is best to avoid tuning priors to previous observations in too substantive a way in order to preserve the robustness of independent verification. Typically then, the term ``non-informative" prior actually refers to a ``weakly informative" prior.

At first glance, one might think that providing an equal weighting to all parameter values ({\em i.e.} uniform in probability) would make the most intuitive sense for the non-informative case. However, this runs into two issues, one trivial and one non-trivial. The trivial issue is that such a prior is improper, not having a finite integral, and also begs the question of whether you actually believe, for example, that it is equally likely to detect 10$^{10}$ events as it is to detect 3. However, in practice, the prior is always multiplied by the likelihood function, which suppresses its impact outside of the region of interest for the actual observation and makes the exact form of the prior far from this region irrelevant. Thus, a uniform prior should really be viewed as a sufficient approximation to one that actually trails off to zero at some point in a manner that does not need to be specifically defined. The non-trivial issue with uniform priors is that uniformity is not necessarily preserved for parameters couched in a different form. Thus, for example, a prior distribution that is uniform in the parameter $S$ is not uniform in $S^2$. This therefore requires a choice to be made as to what form of the model parameters might be reasonably assumed to have a uniform prior probability distribution.

The subject of non-informative priors is one of active discussion and debate. One common alternative is to use a Jeffreys prior~\cite{Jeffreys}, which uses the Fisher information content of the likelihood function itself to define a non-informative prior that is transformationally invariant. While this approach often works well for simple situations, the application of this is not always straightforward for realistic analysis scenarios that may involve multiple, multi-dimensional signal and background components, often with non-parametric forms, nuisance parameters for error propagation, and multi-parameter signal models. Consequently, this can also lead to forms of the prior that are non-intuitive and depend on the particular analysis in which it is applied.

It should be emphasized at this point that there is often no ``correct'' choice of prior. Making any statement regarding the viability of a given model based on the data necessarily requires a stated frame of reference, and the prior just defines the context within which one chooses to make this assessment. With this understanding in mind, we take the pragmatic view that a non-informative prior should be intuitively reasonable, simple to apply and visualize, and must allow for the impact of an alternative choice of prior to be easily evaluated. To this end, we suggest that the use of uniform priors would be preferred and that relevant model parameters should be couched in simple forms that make this a not unreasonable choice. The nature of bounds on different forms can then be derived from this initial determination. This approach has the further advantage that maps of the Bayesian probability densities couched in terms of such ``flat'' parameters are identical to likelihood maps, which can then serve the important dual purpose of providing both a Bayesian assessment of model viability and a detailed presentation of the objective information content of the data.

For the purpose of consistency, it would be advantageous to establish
common conventions ({\it e.g.} PDG guidelines) for such flat parameter forms relevant to different
types of experimental measurement. Fortunately, while the number of
possible models is infinite, the basic nature of fundamental
parameters on which these models depend is not, and we believe that
finding agreement on a set of reasonable choices is not a particularly
contentious issue in practice. As a general rule of thumb, our
perception of model parameters often falls into one of two categories:
either they are variables of magnitude, with values spanning the same
general order and for which a uniform prior (sometimes over a limited
range) may be a reasonable choice; or they are variables of scale,
potentially spanning several orders of magnitude and for which it may
be appropriate to weight such scales equally, resulting in a prior
that is uniform in the log. These two choices typically bound the range
of non-informative priors that are generally considered to be plausible. 
It is, for example, usually difficult to justify a form of non-informative prior that actually rises 
with signal strength or falls faster than would be implied by giving equal weight to all scales.
Examples of
priors uniform in magnitude might include the value of an unknown phase angle, the
value of a spectral index, or the precision measurement of a
quantity whose rough magnitude is constrained. Examples of priors
uniform in scale might include the energy scale for new physics, the cross section for some non-standard
interaction, or first measurements of quantities whose rough magnitude is not constrained. 
A good indication of the relevant variable class can
often be taken from how they are typically represented on parameter
plots ({\em i.e.} whether they have linear or logarithmic scale axes). 

However, for models that directly depend on a counting rate
measurement (especially in the region of low event numbers), and where there is not a strong case for another form of prior, 
we make the pragmatic proposal to always use a prior proportional to a uniform average counting rate. 
This is because the sensitivity range for an
experiment to detect a signal not previously established does not
typically stretch over several orders of magnitude and, in the event
that an upper bound is appropriate, this prior choice produces a
conservative number for evaluating the viability of model parameter values. In addition, as previously shown, 
this choice also tends to yield a good degree of statistical coverage for simple cases
which, while not necessary for Bayesian bounds, we regard as convenient.

\subsection{Sensitivity to Prior}

From Equation \ref{Bayes} it is clear that, for a given hypothesis,
$H$, and data set, $D$, the posterior probability is related to the
prior as follows:

\begin{equation}
\log(P(H|D)) = \log(P(D|H)) + \log(P(H)) + C
\end{equation}
\[= \sum_{i=1}^n \left[\log(f(x_i|H)) + \frac{1}{n}\log(P(H))\right] + C\]

\noindent where $f$ is the likelihood function evaluated for each of $n$ independent data values $x$ and $C$ is a constant. Hence, the impact of the prior probability, $P(H)$, becomes less significant as the number of events increase. However, it is still the case that the choice of prior can, in some instances, have a notable impact on the perception of model viability. The effect becomes particularly relevant where parameter ranges are unconstrained over orders of magnitude and arguments hold for choosing a nominal prior that is uniform in the log. Consequently, while the analysis of the previous section can provide a reasonable basis for default parameter representations and priors, in such cases we believe it is also important to specifically indicate the extent of the prior sensitivity. As a reasonable and pragmatic approach, we suggest doing so by comparing the results from the choice of a prior that is uniform in scale with that which is uniform in the magnitude of the relevant parameter. The impact can be indicated by an additional contour on parameter maps and, if significant, can then be further highlighted in the data summary.

It is frequently the case that the appropriate choice between the two suggested forms of prior is clear and that either conservative upper bounds can be derived in the case that the data is consistent with no signal, or that the strength of an observed signal will help to dictate robust bounds. However, if the choice of prior is not obvious, and if the conclusions strongly depend on the available choice, then avoiding a Bayesian method does not alter this fact. Using purely frequentist approaches will merely hide the ambiguity, potentially leading to false conclusions regarding the robustness of implications for model parameter values. 

When taken together, we believe that the type of approach outlined, involving: 1) a pragmatic choice of prior; 2) an explicit presentation of the likelihood; and 3) a test of the prior sensitivity where appropriate, can provide a robust approach for indicating Bayesian model constraints as an important component of the overall presentation of results. In addition to helping avoid confusion in interpretation, this would also serve the valuable purpose (often overlooked) of explicitly indicating the strength of information in the data and when reliable inferences regarding model parameter values might be made.

\subsection{Unified Likelihood Maps and Data Summaries}

As mentioned in the previous section, the suggestion is to display the
Bayesian posterior probability distribution in terms of model parameters with uniform priors,
which then simultaneously shows the global likelihood as well.
This probability should be suitably marginalized over the non-essential
parameters. It is
sufficient to give the ratio of the Bayesian probability (likelihood) at any one point to the
maximum value. In fact, it would seem sensible to couch this as
$-2\log(L/L_{max})$, since this is often approximately equivalent to differences in
$\chi^2$ from the best fit and, thus,
carries some intuition. This also readily allows for approximate 
frequentist intervals to be inferred via Wilks' Theorem, as discussed in
Appendix A.  
As is often done for 1-parameter models, a
simple graph of this quantity as a function of the parameter can be
shown; for a 2-parameter model, a 2D contour or color map can be
given; and for higher order models, appropriate sample 2D slices
(preferably in the most slowly varying parameters) can be shown.

\newpage

Bayesian credible intervals can be superimposed on top of these
plots and are simply computed by integrating the Bayesian probability distribution 
(in this case equivalent to the distribution of $L/L_{max}$) in the space
of these flat parameters to find the fraction, $f$, of their
distribution above some value $L>L_C$. The contour defined by that
value of $-2\log(L_C/L_{max})$ then corresponds to the credibility
level $CI=f$. This approach also gets around another common problem of
attempting to interpret the meaning of maximum likelihood contours in
terms of significance levels by either relying on Wilks' Theorem
(which, while often providing good estimates, cannot always be relied
upon for precision in the region of low numbers of events and/or near a
physical boundary), or undertaking a potentially burdensome (in some
cases, perhaps intractable) Monte Carlo calculation. By contrast, the
Bayesian calculation is always well-defined and has no such issues.

For the purposes of abstracts, text, tables etc., a convenient summary
of results is desired. In the case of single parameter models, the
value of the parameter corresponding to the maximum likelihood 
(which is equal to the maximum of the posterior distribution) can be
quoted along with the associated Bayesian credible interval. For
the multi-parameter case, results may be summarized by quoting the
marginalized Bayesian intervals for each parameter. As a shorthand for more detailed
likelihood shape information, especially where behavior tends to be
non-Gaussian, we suggest that bounds in terms of flat parameters be
quoted at 2 different credibility levels, for example, 90\% and 99\%,
or 68\% (1$\sigma$) and 95\% (2$\sigma$). And, as previously
discussed, for cases where the choice between the two types of priors
leads to a notably less conservative constraint, the impact should be
specifically indicated.

\subsubsection{Example 1: neutrino mixing}

As a first example, consider the case of neutrino oscillation and mixing. Regarding the choice of non-informative prior for $\Delta$m$^2$, if the scale has not been firmly established, the value could correspond to a range of scales and is generally plotted logarithmically. Accordingly, this suggests the choice of a prior that is uniform in the log of this variable. If the scale has already been well established in the field, a prior uniform in $\Delta$m$^2$ may be more appropriate (though the impact of the choice is unlikely to be substantial in this case). 

For the unknown mixing angles and phases, non-informative priors that are uniform in circular range are suggested. One complication here is that the form the mixing angle takes in, for example, the 2-neutrino vacuum mixing expression is $\sin^2 2\theta$, which means that a given observation lead to an ambiguously defined value of $\theta$ itself, which can be in any of 4 quadrants. Indeed, there is no clear consensus in the field as to the best form to use for such experiments, with various choices including $\sin^2 2\theta$, $\sin^2 \theta$ and $\tan^2\theta$, making comparisons between different results and different phenomenology papers troublesome. While various trigonometric forms may appear to describe different phenomena, the fundamental parameter is the angle itself and it therefore seems appropriate to try to couch things in terms of this variable. The issue of redundant multi-quadrant values that may arise from measurements in some cases can be readily taken into account by using variables such as $|\theta+n\pi|$, which then allows a non-redundant range in theta to represent other quadrants simultaneously.

\begin{figure}
\includegraphics[width=90mm]{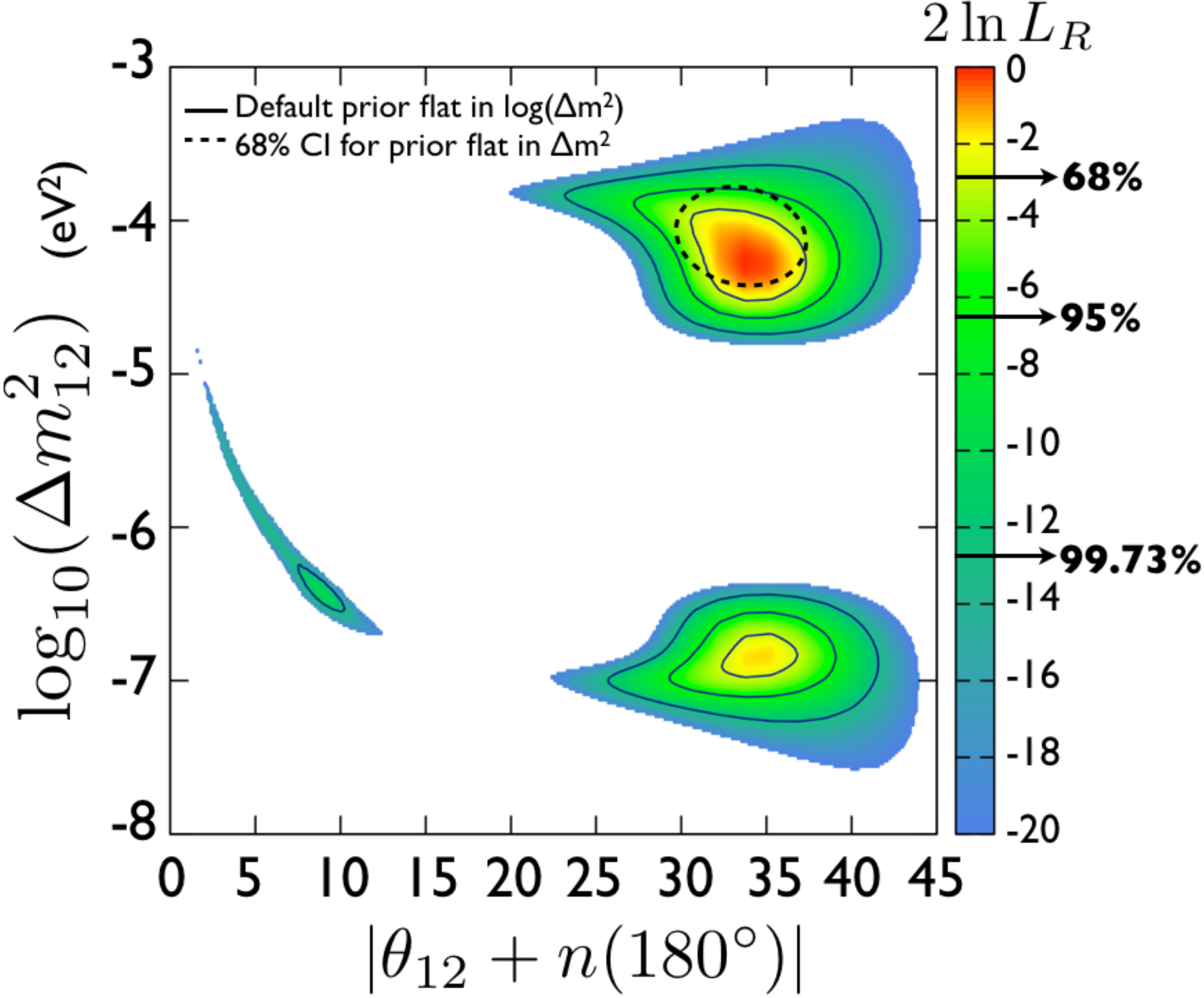}
\caption{Confidence intervals derived from the SNO experiment's salt
  phase data~\cite{sno}.  The colors represent the difference in the log of the likelihood ratio or, equivalently, differences in the posterior
  probability calculated with uniform priors for both axes, relative to
  the maximum point.  The solid black lines are 68\%, 95\%, and
  99.73\% Bayesian contours found by integrating the posterior 
distribution.
  The black dashed region is the 68\% credible region found under an alternative
  prior that is uniform in $\Delta m^2$ rather than $\log(\Delta m^2)$}
\label{SNO}
\end{figure}

As a more specific example of such a construction, we use the publicly available contour map data from the 2005 SNO salt phase solar neutrino mixing parameter analysis~\cite{sno}. At the time of this data, the scale of $\Delta m_{12}^2$ had not yet been unambiguously established by solar neutrino data, so we choose a prior that is uniform in the log of this parameter. We therefore obtain the plot in Figure \ref{SNO}.

The color scale represents the range of $\ln (L/L_{max})$, while
solid line contours are also shown corresponding to Bayesian
credibility levels of 68\%, 95\% and 99.73\%. The values of $-2\ln
(L_C/L_{max})$ corresponding to these levels were found to be 2.82,
6.58 and 12.76, respectively, which are not so far from the naive
Wilks' expectation for critical $\Delta\chi^2$ values corresponding to
a 2D parameter space (2.3, 6.1, 11.8). This suggests that
frequentist constraints would look very similar in this case. To
indicate the sensitivity to the choice of prior, the dashed line
contour indicates the 68\% CI region if a prior that was uniform in
$\Delta m_{12}^2$ itself were used. This region is of a very similar
size to the LMA 68\% contour using the default prior that is uniform in
scale, with a relatively modest displacement of boundary
positions. This indicates that this range is relatively insensitive to
the form of the prior and conclusions regarding the model here are
reasonably robust. However, the LOW region would be eliminated,
suggesting that the default choice of a prior uniform in $\log( \Delta m_{12}^2)$ is the more conservative
approach in this case.

For the data summary, we obtain the marginalized uniform Bayesian intervals $|\theta+n(180^\circ)| = 33.8^{+2.3(4.6)}_{-1.7(4.4)}$ degrees at the 68\%(95\%)CI. For $\Delta m_{12}^2$, we have the interesting situation that the SNO data alone cannot separate LOW and LMA solutions at the desired credibility levels with the assumed prior, giving rise to a bi-modal distribution in the marginalized parameter. In this case, both possibilities should be presented using the 2-part intervals that are naturally produced by selecting the highest probability densities that constitute 68\% and 95\% of the overall distribution, respectively. Thus, we obtain the marginalized uniform Bayesian intervals $-\log_{10}(\Delta m_{12}^2) = 4.3^{+0.1(0.25)}_{-0.3(0.45)}$ eV$^2$ and $6.88^{+0.03(0.17)}_{-0.03(0.23)}$ eV$^2$ at the 68\%(95\%)CI.

\subsubsection{Example 2: rare event search}

Unfortunately, it is not a common enough practice for experiments to provide detailed likelihood information. Consequently, to provide an example of an analysis operating in a much more restricted range of event numbers, we will consider a hypothetical experiment generically searching for rare events. For simplicity, we will assume this to be a counting experiment with negligible systematic uncertainties and a well-defined background expectation, along the lines of the examples considered earlier in this document. 

\begin{figure}
\includegraphics[width=80mm]{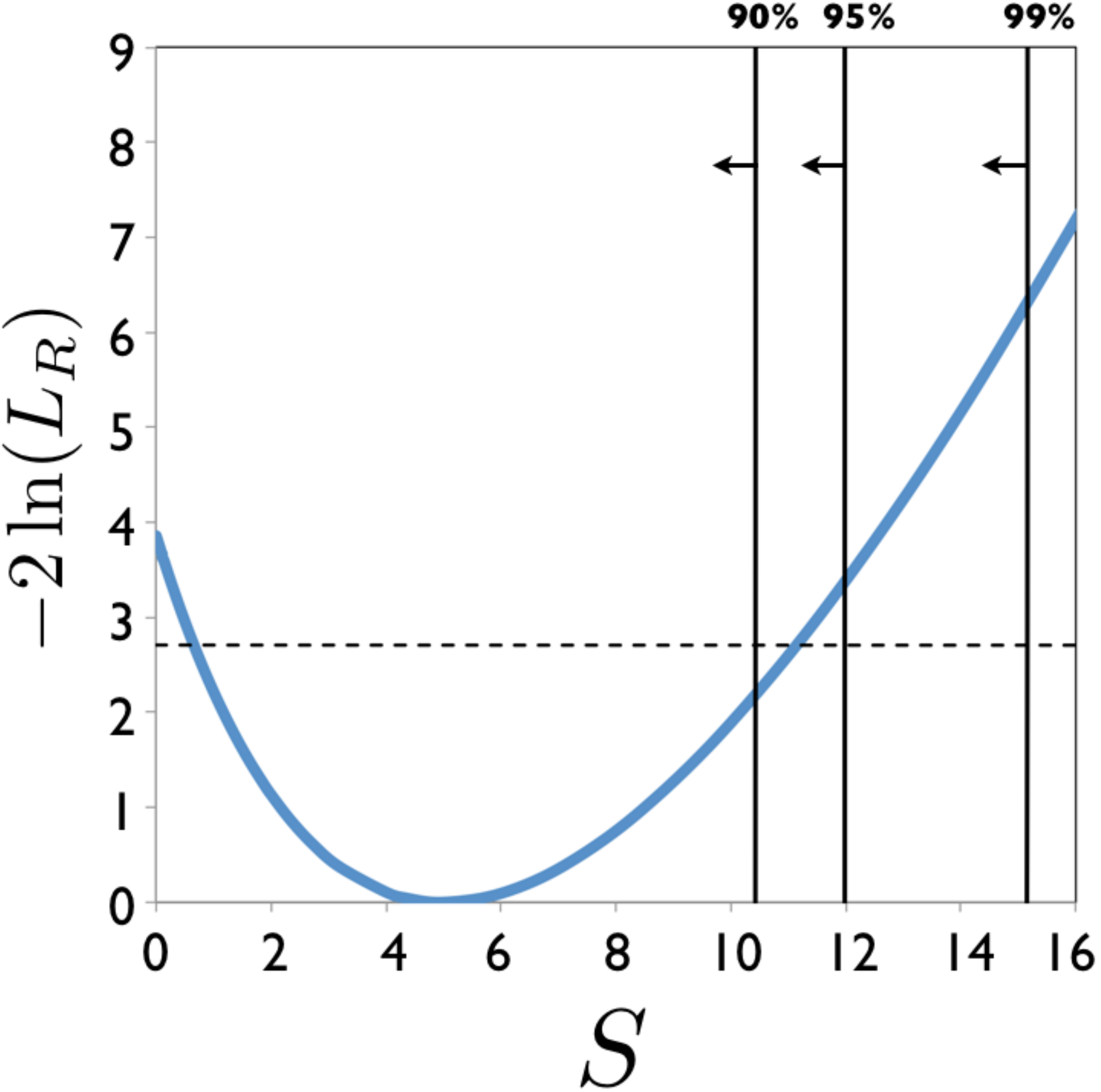}
\caption{Difference in log likelihood from the best-fit signal
  strength, as a function of signal strength, for a Poisson process
  with background rate $B=5$ where 10 events are observed. The
  vertical lines represent Bayesian upper limits at the stated
  confidence levels, while the intersections of the dashed horizontal line
  with the curve indicates the 90\% CL two-sided limits from Wilks' Theorem.}
\label{RareEvent}
\end{figure}

Assume that the expected background is $B=5$ events in the region of
interest and that 10 events are observed. 
We then wish to place 90\%
CI upper bounds on the average signal strength, $S$. In accordance
with previous discussion, as this measurement relates to a low-rate
counting experiment, we choose a prior that is uniform in $S$. We would
therefore use the Poisson likelihood function from Equation
\ref{BayesPoisson} with $B=5$ to plot $-2\ln(L/L_{max})$ as a function
of $S$, as shown in Figure \ref{RareEvent}.

Even for such a simple measurement, the likelihood function will typically not, in fact, be a simple Poisson distribution owing to an imperfect knowledge of the expected background and systematic uncertainties, the effects of which must be taken into account by an appropriate marginalization of the distribution. 

The Bayesian uniform prior upper bounds are indicated on the plot for 90\%, 95\% and 99\% CI. As the alternative choice of a prior uniform in $\log(S)$ would result in a smaller value for the bound ({\em e.g.} $S<7.35$ at 90\% CI), it is not necessary to show the impact of this alternate choice since the chosen prior represents the conservative value. Thus, we may simply quote the Bayesian uniform prior upper bounds on $S$ as 10.4(15.2) at the 90\%(99\%)~CI.

The dashed line also shown on the plot corresponds to an approximate frequentist 90\% CL based on Wilks' theorem, and suggests that an interval would be called for under a unified approach spanning 0.66-11.1 (comparable to the corresponding Feldman-Cousins interval of 1.2-11.5). Hence, for this case, frequentist numbers are reasonably similar to those from the Bayesian approach, with the slight difference at the upper end largely due to the choice to quote a 1-sided versus a 2-sided bound.

\vskip 0.1in

\begin{figure}
\includegraphics[width=80mm]{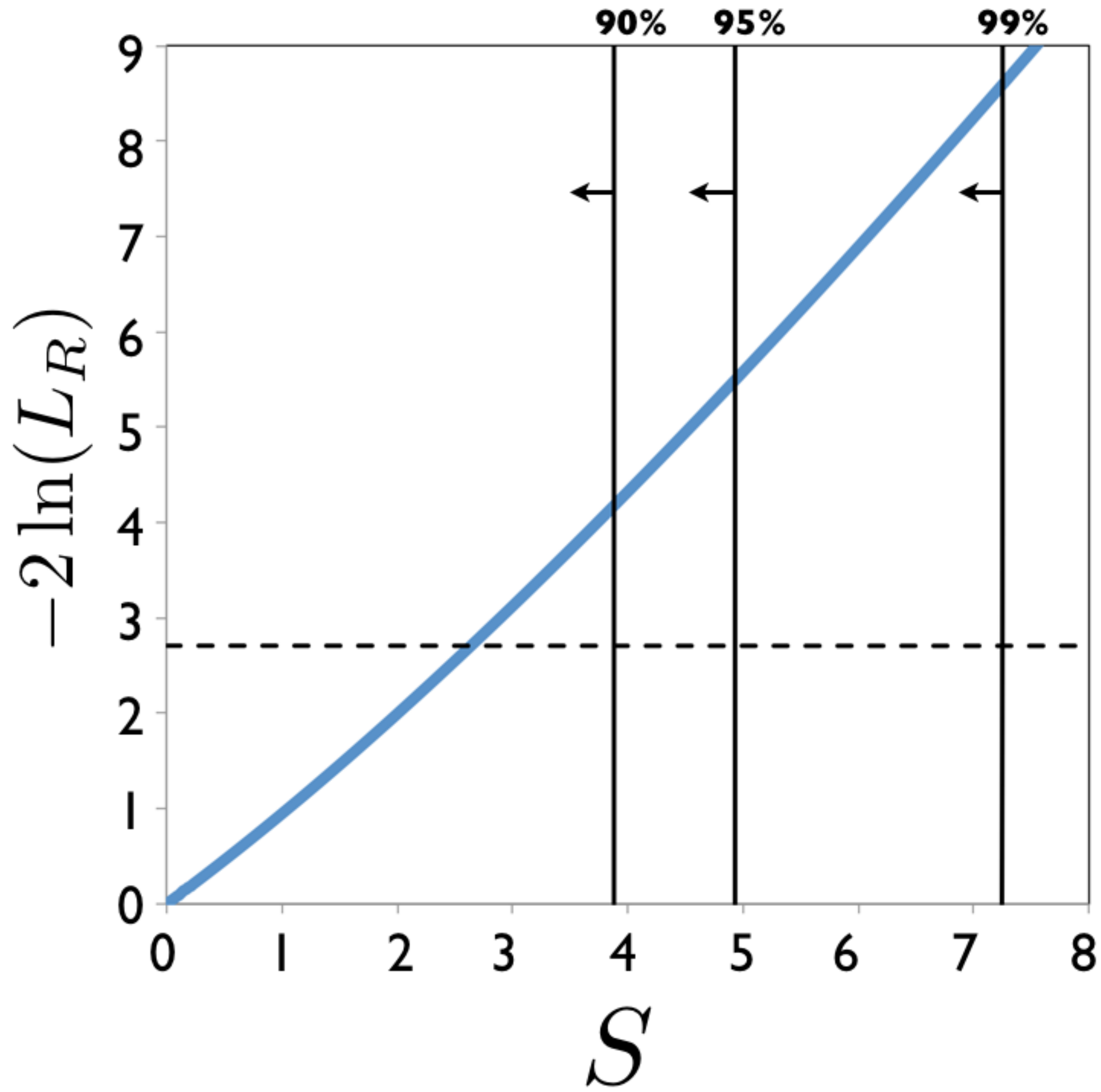}
\caption{Difference in log likelihood between the best-fit signal
  strength, as a function of signal strength, for a Poisson process
  with background rate $B=9$ where 5 events are observed. The
  vertical lines represent Bayesian upper limits at the stated
  credibility levels, while the intersections of the dashed horizontal line
  with the curve indicate the two-sided 90\% CL limits from Wilks' Theorem.}
\label{RareEvent2}
\end{figure}

If we instead considered the scenario described earlier where $B=9$ and 5 events are observed, the likelihood plot shown in Figure \ref{RareEvent2} is instead produced.

In this case, as previously described, the Bayesian upper bound is
3.88, while the frequentist bound inferred from Wilks' theorem is only 2.64
(slightly more conservative than the Feldman-Cousins bound of 2.38 for
this scenario).
This is an example where frequentist bounds are misleadingly
foreshortened as a result of negative fluctuations and prone to
misinterpretation.  

\vskip 1.4in

\section{Conclusion}

The distillation of constraints on theoretical model parameter values is central to scientific enquiry and represents the ultimate goal of experimental observations. There is therefore an understandable desire to interpret the probable nature of such constraints as well as to define the objective details of experimental observations, even when the available information content of the data is necessarily limited. A failure to recognize both aspects and clearly represent them with distinct and appropriate formalisms inevitably leads to misinterpretation and misuse of quoted confidence intervals.

Frequentist confidence intervals, even ``sophisticated''
implementations such as Feldman-Cousins, generally suffer from significant
shortcomings in this regard:
\begin{itemize}
 \item They do not reliably indicate the relevance of individual measurements. 
 \item They are prone to frequent misinterpretation that can result in incorrect conclusions and seemingly paradoxical behaviors.
 \item They do not provide a robust basis for the comparison of different experimental results.
 \item There is no self-consistent approach to propagate systematic uncertainties.
 \item There is no self-consistent method for producing a lower-dimensional confidence interval with a desired statistical coverage
  from a higher-dimensional interval.
 \item They do not possess a unique definition and represent a very limited view of the underlying likelihood information.
 \end{itemize}
Furthermore, frequentist intervals are rarely, if ever, actually used for the purpose in which their context is meaningfully defined: constraining model parameter values at the designated confidence level with a large ensemble of similar measurements. Even where multiple measurements exist, the practice is generally to undertake a combined analysis with either the raw measurements or likelihood maps from individual experiments rather than to compare frequentist intervals. As such, the relevance of quoting intervals that are never actually used for their intended purpose (and, in fact, are readily abandoned with the addition of further data) is questionable. As far as objectively conveying the information content of the data, we therefore believe that providing the likelihood itself as a function of relevant model parameters offers the clearest and most useful approach. 

Bayesian credible intervals offer the only well-defined mathematical approach to placing bounds on model parameter values themselves and, thus, are a necessary component of data presentation if one is to address this aspect. Bayesian intervals are generally free from many of the problems that plague frequentist intervals, but suffer from the one issue of requiring the specification of a prior to establish the context of the model. However, we argue here that 
\begin{enumerate}
\item For Poisson statistics in particular, the use of a prior that is
  uniform in the counting rate produces conservative upper bounds and offers a
  pragmatically defensible choice for bounding model parameter values.
  (Appendix A shows that these credible intervals also have
  reasonable frequentist coverage for simple cases.) 
\item  In other areas, the forms for priors are effectively bounded between those that are uniform in magnitude and those that are uniform in scale ({\em i.e.} equal weighting in the log), where the appropriate choice between these two alternatives for a given parameter class is usually obvious. 
\item When appropriate, explicit indications of the sensitivity to this choice of prior can easily be shown, which provides additional valuable details regarding the strength of information contained in the data that is otherwise hidden (rather than avoided) by pure frequentist methods. 
\end{enumerate}
Taken together, we believe this provides a clear and robust approach to the presentation of model constraints. Achieving a community consensus on such an approach for Bayesian intervals seems both straightforward and of more practical benefit than continuing to only use frequentist constructions which are, themselves, arbitrary and prone to the numerous difficulties previously listed.

We also recommend that the global likelihood for data sets be shown as a function of `flat' Bayesian model parameters to simultaneously indicate the Bayesian posterior probability distribution and objectively represent the relevant information content of the data. Such representations allow other analysts to more optimally combine information from other experiments and test models under different assumptions. Bayesian credibility levels can be readily indicated on such plots and approximate frequentist intervals can also generally be extracted if desired by appealing to Wilks' Theorem (which is found to work well for Poisson statistics, even for small numbers of events). 

As previously indicated, we believe the standardization of priors is straightforward and no more arbitrary than choosing any other type of interval construction.
For data summaries, we therefore recommend quoting Bayesian credibility
bounds in terms of common flat parameter forms, marginalizing where appropriate for multi-dimensional parameter spaces. As a shorthand for more detailed likelihood/Bayesian shape information, we also suggest that, where convenient, such bounds are quoted at two different credibility levels, such as 90\% and 99\%, or 68\% and 95\%. 

Overall, we believe this to represent a much more well-balanced approach to the presentation of experimental results that offers a higher degree of relevance and transparency than purely frequentist conventions, while still providing objective information about the measurement in a more useful form.

\vskip 0.2in

We would like to thank Louis Lyons and Josh Klein for useful discussions and, in particular, Dean Karlen for a careful reading of an earlier draft of the paper. This  work has been supported by the Science and Technology Facilities Council of the United Kingdom and the National Sciences and Engineering Research Council of Canada.  


\appendix

\newpage

\section{APPENDIX A: COVERAGE AND CONSISTENCY}
\label{coverage}

\subsection{Frequentist Coverage}

Central to the frequentist paradigm is the concept of statistical
coverage: that limits derived from an ensemble of repeated experiments
would correctly bound the true model a known fraction of the
time. This is a hypothetical construction, independent of the prospect for actually achieving a large
enough ensemble of measurements to reliably bound a model with the
relevant accuracy this way. Indeed, as there is no clear criteria for when an ensemble is large enough for frequentist intervals to ever be used to constrain model parameter values, nor how this could even be done in principle 
outside of ultimately employing Bayesian statistics, the construction is hypothetical even where there is a large ensemble of measurements.

In this context, the necessity to produce
precise frequentist bounds can seem a little unclear. In fact,
as discussed in Section \ref{nuisance}, 
achieving exact coverage 
is often not actually possible in the presence of systematic uncertainties, and
is not possible for counting statistics even without such uncertainties, 
since the quantized nature of measurements inevitably leads to
coverages that are either a little greater than the target confidence
level (over-coverage) or less than this target (under-coverage). The
former of these is generally chosen to insure coverage in excess of the target confidence level
(though one might pragmatically
argue that such intervals need only be determined to an accuracy
comparable to that with which they will ever actually be used to bound
a model). Consequently, it is possible to spend large amounts
of CPU time computing Feldman-Cousins intervals in which the achieved
precision  
does not reflect the achieved accuracy.

\begin{figure}
\includegraphics[width=85mm]{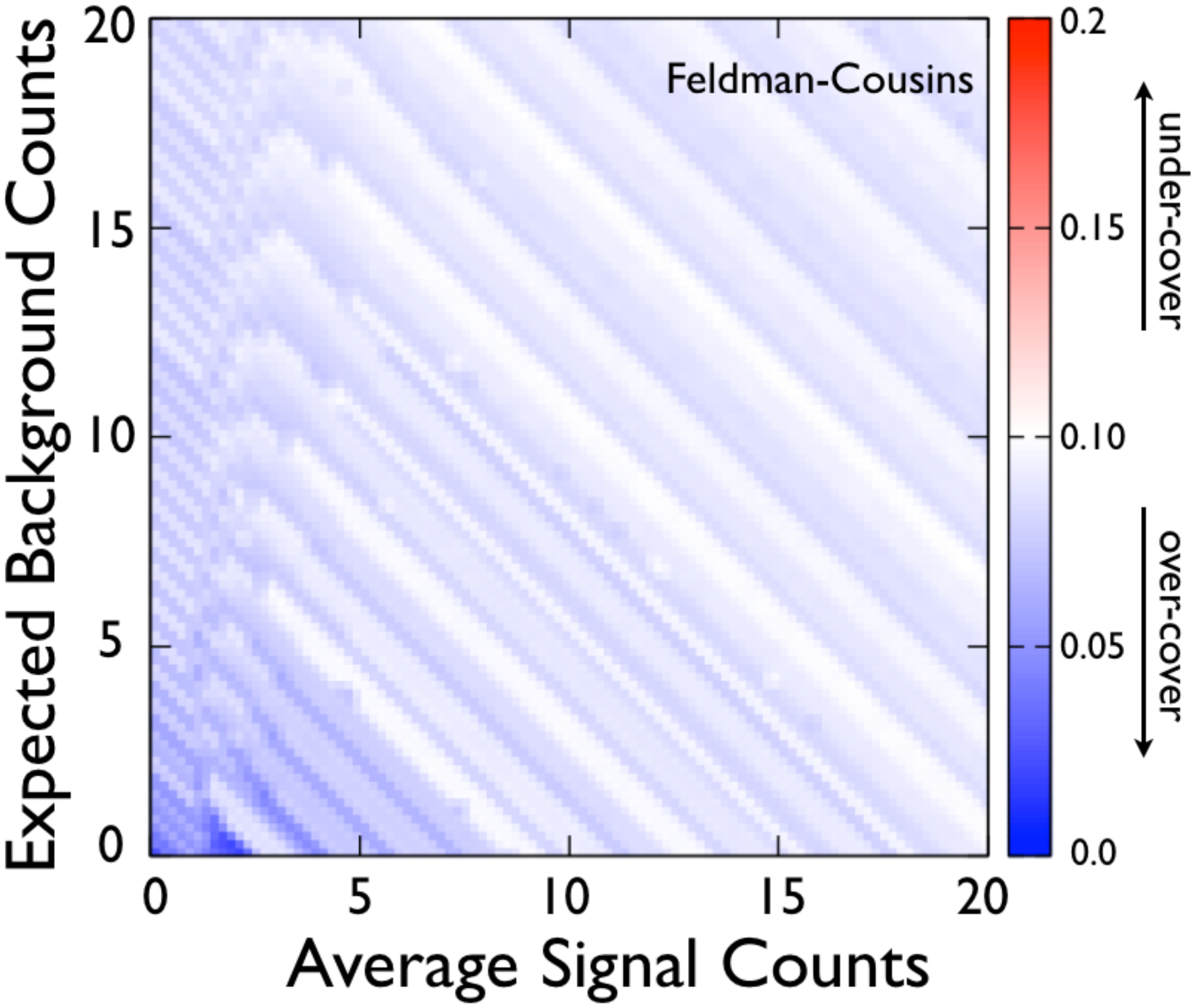}
\caption{Fraction of time with which a Feldman-Cousins 90\% confidence
  interval generated from Poisson observations does not contain the true
  signal rate, as a function of the true signal rate and the
  background rate (which is assumed to be known perfectly). The
  over-coverage of a few percent is intrinsic to the discrete nature of
  the observable.}
\label{FC_Map}
\end{figure}

The coverage map for the Neyman construction of 90\% CL intervals
using the Feldman-Cousins ordering rule is shown in Figure
\ref{FC_Map} for Poisson statistics in the region of low counts. This figure
plots the frequency with which the true model parameter value lies
outside of the derived bounds and indicates the extent and
distribution of the inherent over-coverage. This assumes a perfect
knowledge of the background level, no systematic uncertainties, etc.
Here the over-coverage, which can be as large as a few percent for
certain values of the signal and background rates, is due entirely to
the discrete nature of Poisson counting statistics.

\begin{figure}
\includegraphics[width=85mm]{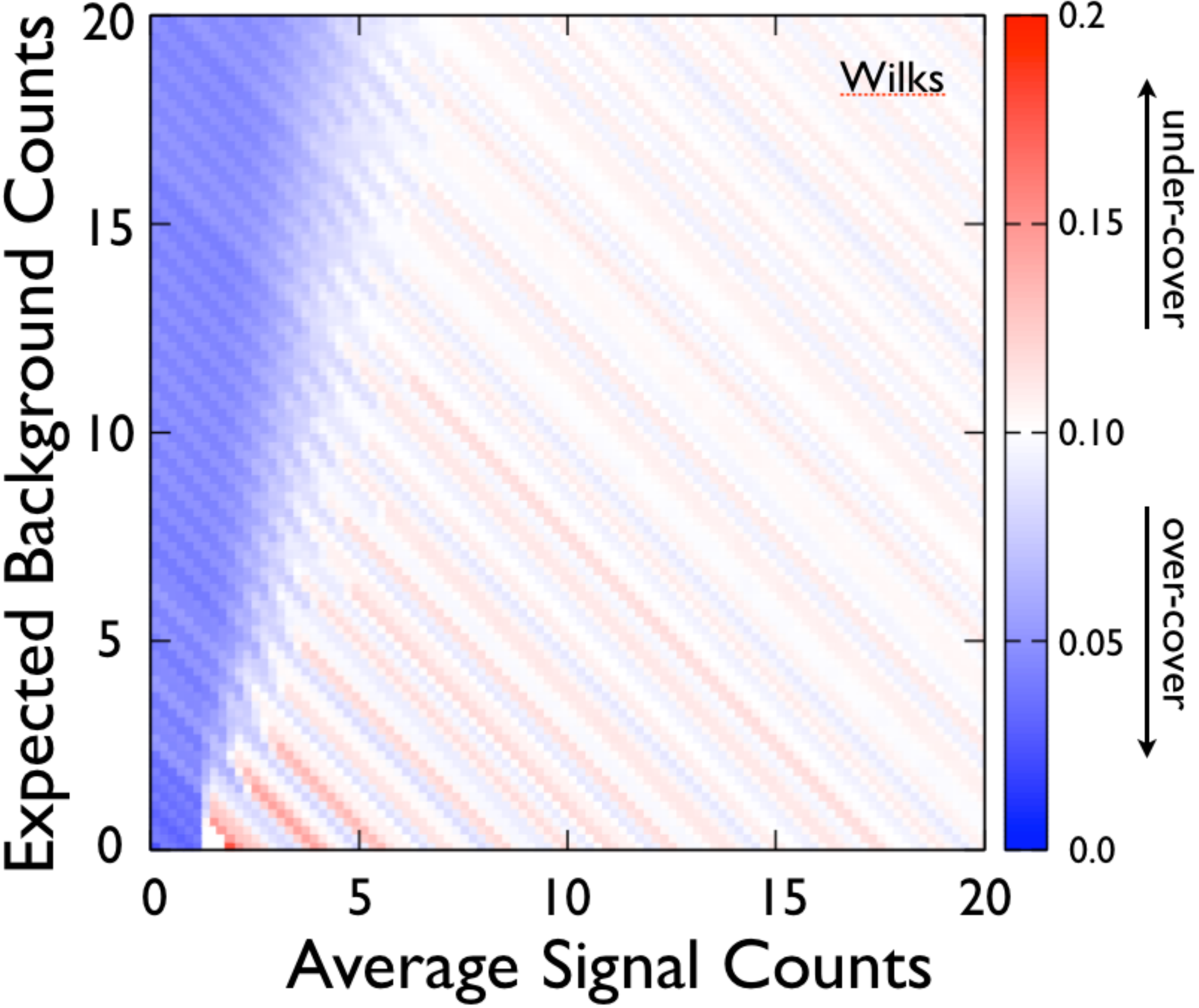}
\caption{Fraction of time with which 90\% confidence intervals generated by Wilks' theorem for a Poisson observation do not contain the true  signal rate, as a function of the true signal rate and the background rate (which is assumed to be known perfectly). The magnitude of deviations from the expected coverage are comparable to the Feldman-Cousins case.}
\label{Wilks_Map}
\end{figure}

While the Neyman construction permits the most controlled definition
of frequentist intervals, it is also often usefully the case that
approximate frequentist intervals can be much more easily derived from
the global likelihood itself by appealing to Wilks' Theorem, which
says that, in the large $n$ limit, $-2 \log L_R$ is distributed as a
$\chi_d^2$ distribution, where $L_R$ is the likelihood ratio between
nested hypotheses defined by $d$ model parameters. In fact, this works
surprisingly well for Poisson statistics, even for small numbers of
counts (as might be implied by the fact that the Bartlett correction
for this case is estimated to be small~\cite{bartlett}). The corresponding 
coverage map for this is shown in Figure \ref{Wilks_Map}, where a $\chi^2$ threshold 
of 2.706 is used for $-2 \log (L/L_{best})$ to define a 2-sided 90\% CL interval similar to Feldman-Cousins, 
with $L_{best}$ evaluated for non-negative values of $S$ (which leads to the
over-coverage seen on the left side of the plot). A very good
approximation to 90\% coverage is achieved and, in fact, for most of
the plane, the magnitude of the deviation from exact coverage is no larger 
than that seen in the Feldman-Cousins case of
Figure~\ref{FC_Map}.  
 If desired, the areas of
slight under-coverage can be all but eliminated to achieve more
conservative bounds by simply choosing slightly higher threshold
values for the likelihood ratio. 

Conveniently, the accuracy with which
Wilks' theorem approximates the coverage suggests that
reasonable estimates of frequentist intervals can often be given, even for small statistics, by simply providing the global likelihood as
a function of relevant model parameters without resorting to the
computationally intensive Feldman-Cousins construction. 
In cases where the applicability of Wilks' theorem may be more questionable, Monte Carlo simulations
can be used to spot-check the implied coverage in the region of interest.

\vskip 0.1in

\subsection{Bayesian Consistency}
Bayesian credibility levels represent the degree of belief
that the limits derived from a particular observation actually bound
the true value of the model parameter. Unlike frequentist intervals,
the extent of Bayesian intervals are intended to directly represent
the probable range of the model parameter. To make this concept
tangible in the sense of calculation, we define ``Bayesian consistency'' as the 
notion that, if possible model parameter values were sampled according to the assumed
prior distribution and instances were selected where an experimental
measurement would result in exactly what was observed, then the
derived credible intervals correctly bound the true model in the
desired fraction of these cases. And, indeed, correctly constructed Bayesian intervals are
{\em always} exact in this sense for the assumed paradigm, even for the case of
Poisson statistics. As such, unlike frequentist coverage, maps of Bayesian consistency are not
necessary. It is also perhaps worth noting that, since Bayesian
credible intervals are specific to each particular measurement and
have a meaning that is independent of any ensemble, the choice of interval type
({\em e.g.}``flip-flopping,'' 
as discussed in Section \ref{unification}) 
has no impact at all on Bayesian consistency.  

\subsection{Some Unfair Comparisons}

\begin{figure}
\includegraphics[width=85mm]{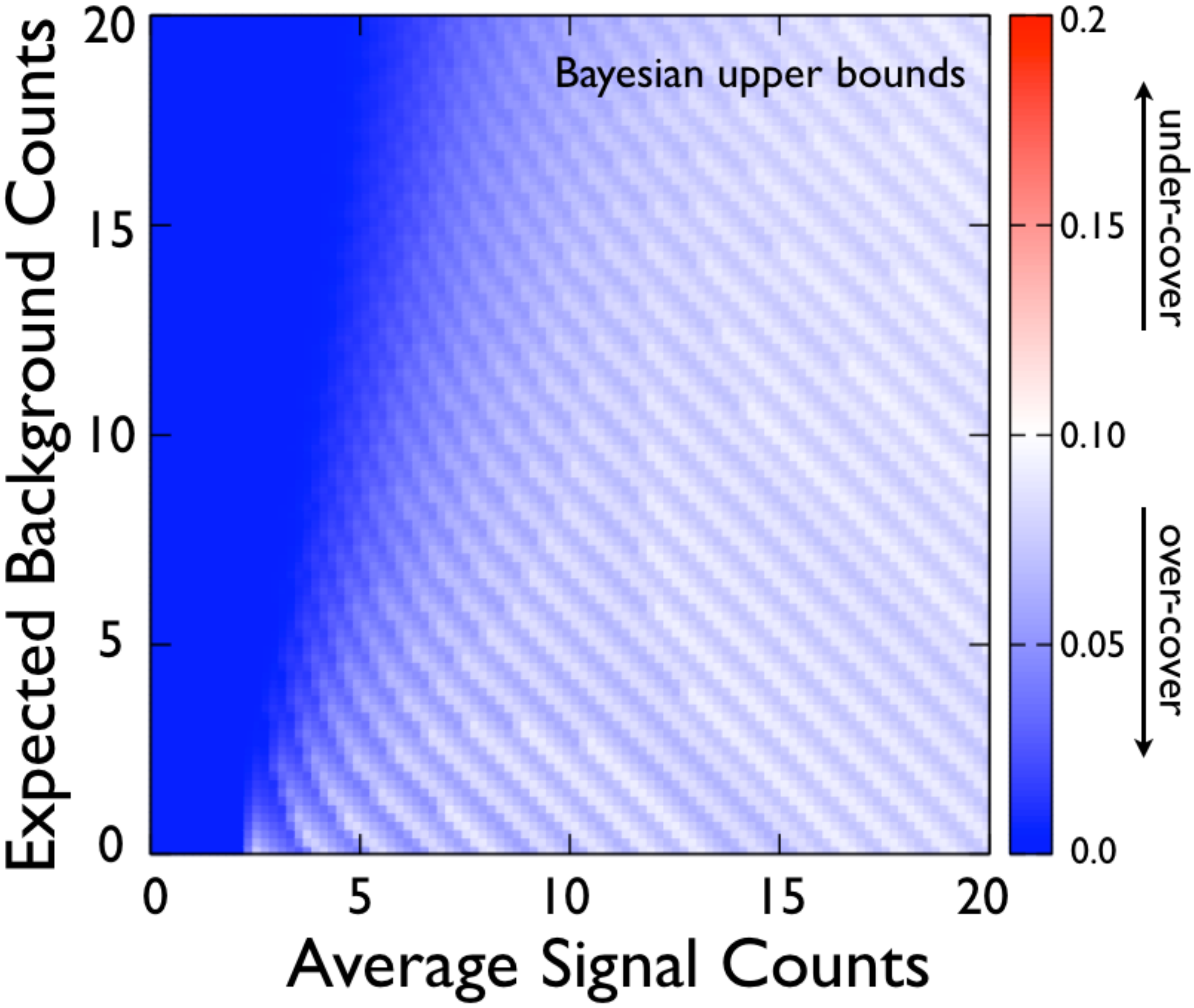}
\caption{Fraction of time with which 90\% CI Bayesian upper bounds
  generated with Bayes theorem and a uniform prior for a
  Poisson observation, and then treated as a CL, do
  not contain the true 
  signal rate, as a function of the true signal rate and the
  background rate.  There is over-coverage, which implies that the
  Bayesian intervals are more conservative than Neyman-constructed
  intervals for this choice of prior.}
\label{BayesUp_Map}
\end{figure}

While never intended for this purpose, it is nevertheless of interest
to explore the extent to which Bayesian constructions also provide
statistical coverage. In particular, we will
explore the case in which a uniform prior in signal rate is used for a
simple, 1-D Poisson observable. Results are shown in Figure \ref{BayesUp_Map} for
Bayesian upper bounds. As can be seen, these bounds tend to
over-cover. In other words, for an ensemble of repeated experiments,
upper bounds constructed using this Bayesian prescription will tend to
contain the true model value with a higher frequency than,
for example, Feldman-Cousins bounds. Therefore, at least for upper
bounds on counting statistics (which is also relevant to estimates of
experimental sensitivities), this prescription is conservative and
sufficient for both definitions of coverage.

\vskip 0.1in
\vskip 0.1in

\begin{figure}
\includegraphics[width=85mm]{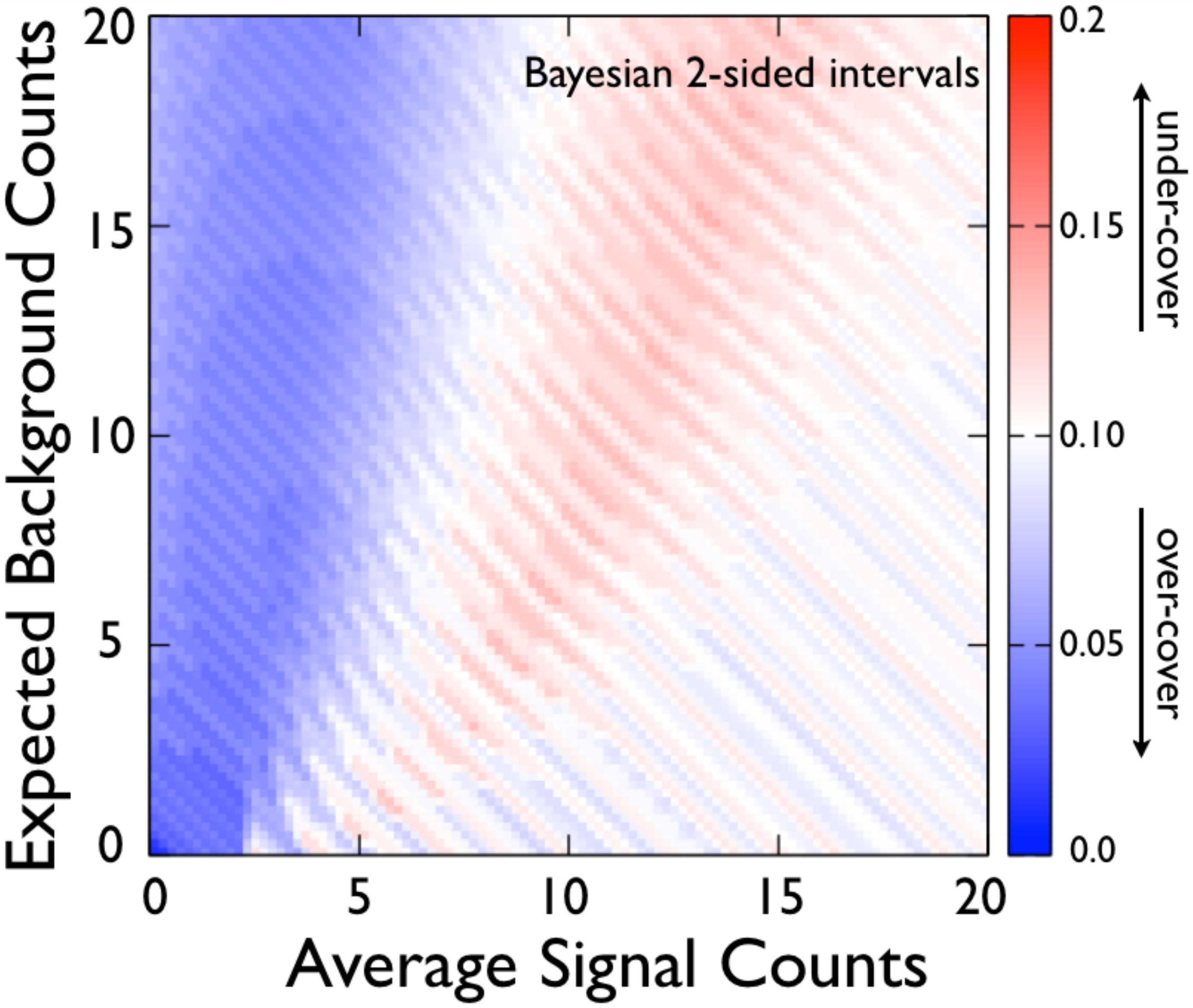}
\caption{Fraction of time with which two-sided 90\% Bayesian credible intervals
  generated with Bayes theorem and a uniform prior for a Poisson
  observation do not contain the true signal rate, as a function of
  the true signal rate and the background rate.  There is mild
  over-coverage or under-coverage, with deviations from the exact
  coverage of approximately the same size as those seen for
  Feldman-Cousins intervals.}
\label{Bayes2_Map}
\end{figure}

For 2-sided interval constructions, the coverage is modified, as shown
in Figure \ref{Bayes2_Map}. Reasonable statistical coverage is
generally achieved in this case as well, although there is some
under-coverage in the region corresponding to borderline signal
detections. For such cases, the 90\% CI
Bayesian bounds provide statistical coverage at the level of 86\% or
more if treated as a frequentist CL.

\begin{figure}
\includegraphics[width=85mm]{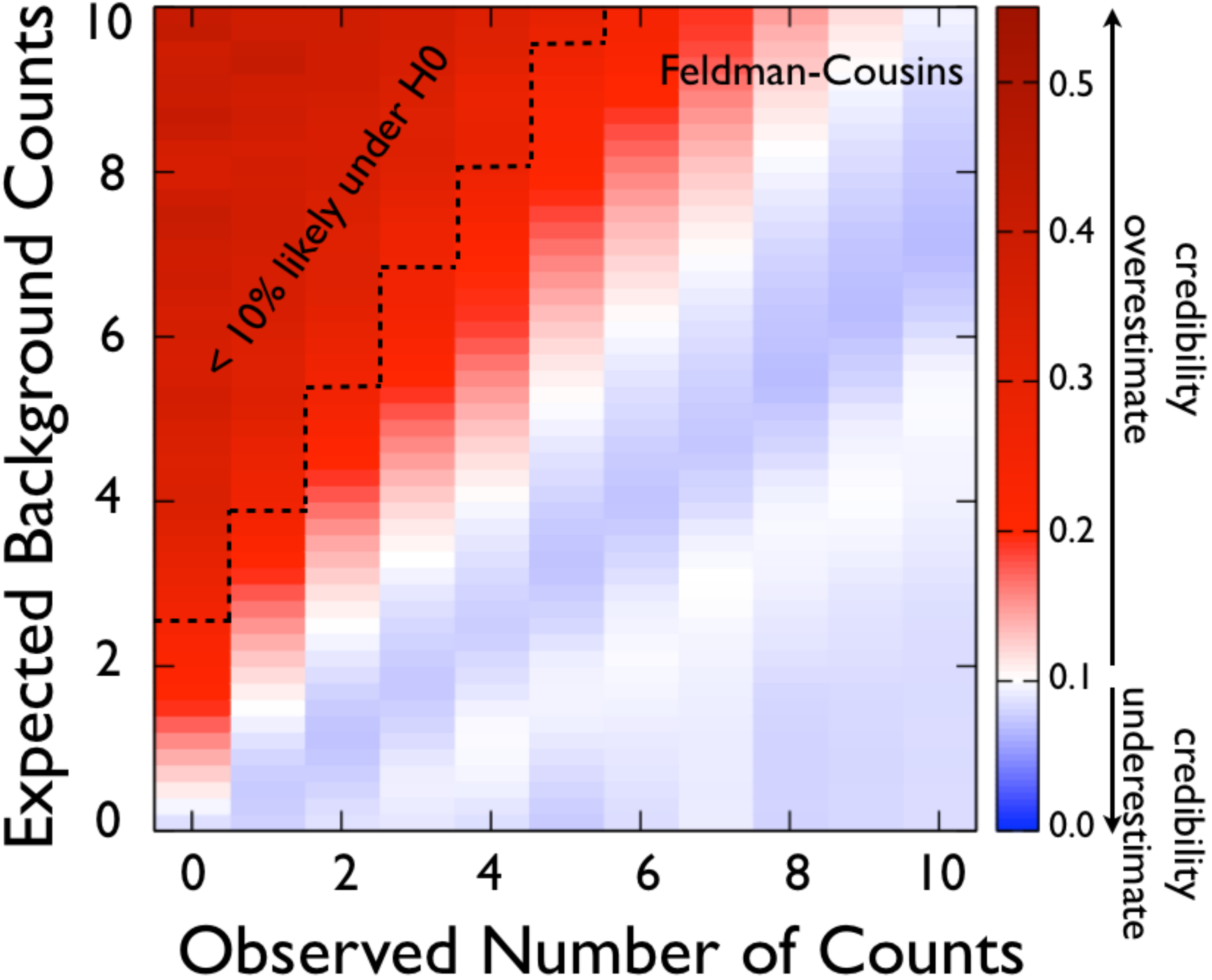}
\caption{Probability that a 90\% CL Feldman-Cousins interval excludes the true mean signal rate given
a particular observed number of counts and expected background rate, where the probability is an
average over equally weighted values of the mean signal rate.
}
\label{FCModels_Map}
\end{figure}

Correspondingly, it is also of interest to explore the extent to which
frequentist constructions also provide Bayesian consistency for this
same case. Figure \ref{FCModels_Map} shows this for the Feldman-Cousins
method and indicates the probability for the true model
parameter value to lie outside of the derived bounds for a given
measurement, assuming that all signal strengths are equally
likely. This plot has a very different interpretation  from Figures \ref{FC_Map}-\ref{Bayes2_Map}: it indicates what 
fraction of true values of the signal rate are not contained in the
Feldman-Cousins interval {\em for a particular observed number of counts},
where each model value is given equal weight in its assigned prior probability.  Here,
the frequentist CL value, treated as a CI, can significantly overestimate the credibility for downward
fluctuations in the observed rate.  The entire upper left portion of
the plot represents cases where the Feldman-Cousins intervals very
often do not contain the true signal rate.  
This again highlights the fact that frequentist intervals only yield correct statistical 
coverage as an ensemble average. However, for particular observations, the 
constructed interval can often be quite unlikely to contain the true value of the parameter. 
While some sections of the upper left region
of this plot represent unlikely fluctuations, there is a reasonable
fraction of phase space where this is not the case. In fact, as the
average signal strength approaches zero, the frequency with which
overestimation of credibility levels occurs approaches 50\%, with significant overestimation
occurring in more than 10\% of cases. This gives rise to many of the
apparent paradoxes previously described when frequentist intervals are
misinterpreted as providing betting odds for parameter values. Hence,
while the Bayesian construction described here also typically provides
reasonable frequentist coverage, frequentist bounds do not generally
provide good estimates of Bayesian credibility.

It may be objected that these conclusions are based on a particular
choice of prior and that  details will change for other
choices. Nevertheless, it is true to say that frequentist intervals
will generally not provide adequate Bayesian credibility estimates for other common
prior choices either. At the same time, while the situation becomes more complicated with multiple dimensions, this example demonstrates it is possible to make a pragmatic choice of a commonly used prior for Bayesian intervals that will also provide a reasonable level of
frequentist coverage for the case of a simple Poisson observable.  While such coverage is in no way required for Bayesian intervals, we regard this property for the simple Poisson case as convenient.

\clearpage
\newpage

\section{APPENDIX B: MAXIMIZATION VERSUS INTEGRATION OVER NUISANCE PARAMETERS}
\label{marginalize}

A point that is sometimes a source of confusion is that the
Bayesian and frequentist approaches for incorporating the effect of
nuisance parameters are often different. Given a joint likelihood
distribution $L(\theta,\xi)$, a 1D Bayesian probability distribution
for $\theta$, assuming a uniform prior $\pi(\xi)$ for $\xi$, is given by:
\begin{equation}
P(\theta) \propto \int d\xi L(\theta,\xi) \pi(\xi) \propto \int d\xi
L(\theta,\xi) \equiv L_{marg}(\theta) 
\label{eq:integ}
\end{equation}
where we may refer to $L_{marg}(\theta)$ as the ``marginalized
likelihood'' obtained from integrating $L(\theta,\xi)$ with a flat
prior for $\xi$.  (Strictly speaking $L_{marg}(\theta)$ is not a
likelihood function but a posterior distribution, but since it plays
an analogous role in Bayesian analyses to the frequentist profiled
likelihood the notation $L_{marg}$ is useful.)  In contrast, the
frequentist approach to eliminating a nuisance parameter is often to
instead ``profile'' the likelihood:
\begin{equation}
L_{prof}(\theta) = \max_\xi L(\theta,\xi)  
\end{equation}
It is not necessarily intuitive how these seemingly different methods
are related, 
if at all.

In fact, the marginalization by integration given in Equation~\ref{eq:integ} is the more fundamental method, and follows
directly from the laws of conditional probability.  It is not often appreciated that the ``profiling'' method is actually just a
numerical approximation to the full integration.  This approximation, proposed by Laplace \cite{Laplace}, can be derived by rewriting Equation~\ref{eq:integ}:
\begin{equation}
L_{marg}(\theta) = \int d\xi L(\theta,\xi) = \int d\xi \exp ( \ln L(\theta,\xi) ).
\end{equation}
For fixed $\theta$, one then proceeds by finding the maximum of $\ln
L$ as a function of $\xi$ and doing a Taylor expansion around the
value $\hat{\xi}_\theta$ that maximizes $\ln L(\theta,\xi)$: 

\begin{equation}
L_{marg}(\theta) \approx 
\end{equation}
\[\int d\xi \exp \left( \ln L(\theta,\hat{\xi}_\theta) - \\
\frac{1}{2} \left| \frac{d^2 (\ln L(\theta,\xi))}{d\xi^2} \right|_{\hat{\xi}_\theta} (\xi - \hat{\xi}_\theta)^2
\right)\]

To the extent that higher order terms can be neglected, this integral evaluates to: 
\begin{equation}
L_{marg}(\theta) \propto \sqrt{\frac{1}{\left| \frac{d^2 (\ln L(\theta,\xi)}{d\xi^2} \right|_{\hat{\xi}_\theta}}}
 \exp \left( \ln L(\theta,\hat{\xi}_\theta) 
\right)
\end{equation}

This is nothing other than $L_{prof}$ times a correction factor which depends on the second derivative of the likelihood function.  If the likelihood function is Gaussian, the correction factor is a constant and so the approximation is exact, and if the likelihood function is only approximately Gaussian, then its second derivative still only varies slowly with $\theta$.  Expressed more simply in
terms of the log likelihood, we have:
\begin{equation}
-\ln L_{marg}(\theta) \approx \label{eq:laplace}
\end{equation}
\[ -\ln L_{prof}(\theta) + 
\frac{1}{2} \ln
\left(
{\left| \frac{d^2 (\ln L(\theta,\xi)}{d\xi^2} \right|_{\hat{\xi}_\theta}}
\right)  
\]
Hence, we see that the frequentist's method of eliminating nuisance parameters by maximizing the likelihood as a function of the unwanted parameter is actually just an approximation to integrating over the parameter in the case that the joint likelihood is Gaussian.  Furthermore, Equation~\ref{eq:laplace}  gives us a correction factor to this approximation that can improve the accuracy in the case that the likelihood is not exactly Gaussian.

To the extent that the likelihood function is well-approximated by a multi-dimensional Gaussian, even a Bayesian analysis may find the multi-dimensional generalisation of
Equation~\ref{eq:laplace} a useful numerical expedient to replace a potentially complicated integral with a function minimization and the evaluation of the second derivatives of $\ln L$ at that minimum.

\newpage

\section{APPENDIX C: MAXIMUM SIZE OF FREQUENTISTS "FLIP-FLOP" EFFECT}
\label{flipflop}

The ``flip-flop'' effect is a phenomenon that only impacts 
frequentist
intervals and can occur when different experimenters choose for
themselves when to quote a given type of interval based on the result,
leading to a small statistical bias in frequentist coverage. 
To estimate the maximum size of the effect, consider a signal
with a strength corresponding to an average detected significance
level that is just at the threshold for an experimenter to rule out
the zero signal hypothesis. Accordingly, about half the time, a fluctuation
will take place in the positive direction that will result in a
claimed detection and a 2-sided CL interval. In this case, 
since the observed rate is larger than the true average signal rate, 
a large enough upwards fluctuation can result in an interval with a lower
limit larger than the average rate, which would then incorrectly exclude this value.
For a symmetric confidence interval, the frequency with which such values are
nominally excluded in the lower half of the distribution is
$\frac{1}{2}(1-CL)/0.5 = 1-CL$ (normalizing for only the lower
half). However, when a negative fluctuation occurs, the significance
drops below threshold and a 1-sided upper limit is set. The average
signal is now larger than implied by the observed number of events,
and risks being excluded at the upper end. The frequency with which
signals are nominally excluded in the upper half of such a 1-sided
distribution is $(1-CL)/0.5 = 2(1-CL)$. Therefore, the total exclusion
frequency for positive and negative fluctuations is $\frac{1}{2}(1-CL)
+ \frac{1}{2} [2(1-CL)] = \frac{3}{2}(1-CL)$. 

Thus, the maximum effect would mean that a 90\% CL only has 85\% coverage, and a 99\% CL only has 98.5\% coverage.

\end{document}